\documentclass{jfm}
\makeatletter
\def\absfooterflag{}
\def\pagelimitfooter{}
\def\ps@titlepage{\leftskip\z@\let\@mkboth\@gobbletwo\vfuzz=5\p@
  \def\@oddhead{\hfil\llap{\thepage}}%
  \def\@evenhead{\hfil\llap{\thepage}}%
  \def\@oddfoot{}\def\@evenfoot{}%
  \def\sectionmark##1{}\def\subsectionmark##1{}}
\makeatother
\usepackage{graphicx}
\usepackage{epstopdf,epsfig}
\usepackage{mathtools}
\usepackage{newtxtext}
\usepackage{newtxmath}
\usepackage{natbib}
\usepackage{hyperref}
\hypersetup{
    colorlinks = true,
    urlcolor   = blue,
    citecolor  = black,
}

\usepackage{pgfplots}
\usepackage{xcolor}
\usepackage{multirow}
\usepackage{tikz}
\definecolor{ao(english)}{rgb}{0.0, 0.5, 0.0}

\usepackage[caption=false,justification=centering]{subfig}
\usepackage{amsmath,array}
\usepackage{bm}

\title{Particle dynamics in a microfluidic T-junction vortex flow: the role of history force}

\author{Matteo Chiatto\aff{1}
  \corresp{\email{matteo.chiatto@unina.it}}, Luigi de Luca\aff{1}, Howard A. Stone\aff{2}, Francesco Grasso\aff{3} }

\affiliation{
\aff{1} Department of Industrial Engineering, University of Naples ``Federico II'', Naples, 80125, Italy
\aff{2}Department of Mechanical and Aerospace Engineering, Princeton University, Princeton, New Jersey 08544, USA
\aff{3}DynFluid Laboratory, CNAM-Arts et Metiers Paris Tech, Paris 75013, France
}

\begin{document}
	\maketitle
	
\begin{abstract}
The motion of suspended inertial particles in a carrier fluid together with the dynamics of the underlying vortical structures through a square microfluidic T-junction channel are investigated. In particle-laden flows, the coherent structures forming near the branching strongly affect the particle distribution and may promote particle trapping around the T-region. Single-phase numerical simulations of the unperturbed base flow and integration of the particle force balance equation are carried out to characterize the particle motion along the trajectories. Unlike most previous studies, the force balance includes the Basset--Boussinesq (BB) history term contribution. The flow topology is analysed for two Reynolds numbers, $Re$ equal to $500$ and $570$, while particle properties are varied through the Stokes number, $St$, and the particle-to-fluid density ratio, $\rho$. For each $Re$, it is shown how the pair $(\rho, St)$ controls particle accumulation and possible capturing around the T-junction region.
Consistently with previous studies, at $Re=500$ the base flow exhibits a steady vortex pair, whereas at $Re=570$ a Hopf bifurcation gives rise to a helical oscillation of the primary vortex pair characterized by periodic breathing. In the latter case, particle trajectories depend strongly on both injection location and injection time.
The current study shows that the Basset--Boussinesq history force has a non-negligible impact throughout the whole range of $(\rho, St)$ pairs investigated. Analysis of the force balance equation shows that the BB contribution acts as an additional viscous force which, together with the drag of the relative motion, opposes the pressure-gradient force, making particle spiral trajectories wider in the radial direction and favouring its elongation in the axial direction. Inclusion of BB contribution reduces the basin of attraction (BoA), indicating that memory effects promote the particle displacement away from the vortex core and weaken the trapping over the whole parameter range.
At $Re=570$ a modal analysis is used to characterize the frequency content of the breathing motion of the primary vortex pair. For a given injection location, particles interact with the vortical structure at different phases of the oscillation cycle depending on the injection time, thereby producing substantially different trajectories. The resulting phase-dependent dynamics is reflected in the shape and extent of the BoA and is quantified in terms of the trapping probability over the injection inlet section, and Probability Density Function (PDF) of the various force components as well.
\end{abstract}

\section{Introduction}\label{sec:intro}

Flow configurations typical of pipeline networks that transport working fluids like  water, oil and chemicals, or biological systems such as the cardiovascular network for blood distribution, involve bifurcations in the geometry that divide the incoming flow into two or more branches. The flow near a branching geometry may evolve into secondary structures, that usually consist of recirculation regions or complex vortical zones \citep{ Vigolo2013236, Vigolo20144770, Kim2021}. Secondary structures, on the one hand, can introduce undesired pressure drops or blockage effects, and on the other, can also be harnessed to introduce specific effects to be exploited as control properties, offering a fascinating avenue to develop innovative devices. Interesting applications are droplets generators \citep{Chen2017401}, heat exchangers \citep{Sadique2022}, and micro-devices for size-selective collection of tumour cells \citep{Sollier201463}. The dynamics of solid  and low-density particles suspended in fluid flows has also gained great attention \citep{Uspal2013, Magnaudet2000659, Zhang2021}.

A configuration studied both theoretically and experimentally is the T-junction geometry with a square cross-section, where the incoming flow divides in two branches with identical sections. In this context, \citet{Vigolo2013236} developed a simplified model to forecast whether a particle entering a T-junction will collide with the channel wall and to determine the rate at which any damage occurs. In a subsequent work, \citet{Vigolo20144770} observed the trapping of particles around the branching region where the curvature of the streamlines produces centrifugal forces giving rise to a large vortex pair. As a result, above a critical Reynolds number low-density particles or bubbles can be trapped at the junction, leading to particle accumulation and changes in the flow topology. The Reynolds number (defined as $Re=UL/\nu$, where $U$ represents the average inlet velocity, $L$ is the square length and $\nu$ the fluid kinematic viscosity) is a control parameter. According to  \citet{Chen2015} at $Re\geq200$ the vortex pair is well defined and dominates the flow field. At higher $Re$ ($\approx315$), axisymmetric bubble-type vortex breakdowns appear at four locations in the T-junction, producing recirculation bubbles and stagnation points. Interestingly, the streamlines inside the recirculation region first exhibit a converging spiralling motion towards the stagnation point and then a backward motion directed towards the vortex core. A further increase in the Reynolds number produces the occurrence of the first Hopf bifurcation at $Re=556$, where the flow exhibits a helical oscillation of the outlet vortex pair, with the vortex structures corkscrewing and aligning in the streamwise direction \citep{Chen2017257}. 

The junction angle, namely the angle between the incoming flow and the direction of the branches (the T-junction corresponds to a $90^{\circ}$ junction angle), strongly influences the particle capture mechanism. Its effect was investigated both experimentally and through single-phase numerical simulations
by \citet{Ault2016}, by injecting particles of different sizes and densities and considering various arrow- and Y-configurations at different $Re$. Those authors found that the minimum numerical Reynolds number for the onset of the vortex breakdown is associated with the experimental minimum $Re$ for which particles are trapped. This suggests that the vortex breakdown is responsible for particle capture and that single-phase numerical simulations can be used to predict the behaviour of two-phase systems. A further investigation of trapping was carried out by \citet{Oettinger2018}, who identified the trapping region as the largest flow sub-domain where particles are released and remain there for all times. The analysis highlighted that trapping regions takes on an anchor-like shape, with an extension larger than the recirculation regions. More recently, \citet{Chan2018} developed an experimental methodology to make a direct observation of the vortex breakdown process, while the effects of non-Newtonian characteristics on the vortical structures evolution was investigated by \citet{Kim2021}. 

The dynamics of particle motion in an unsteady, non-uniform flow relies on the balance of forces acting on an individual particle. Since the works of \citet{Corrsin1956114}, \citet{Maxey1983883}, and \citet{Gatignol}, the form of the balance equation has been progressively refined through improved modelling of the different force contributions. Particles, modelled as rigid spheres, experience the action of fluid acceleration, buoyancy, viscous drag of the particle-fluid relative motion, lift and the augmented viscous drag associated with the Basset–Boussinesq (BB) history (or memory) term. In particular, the BB term takes into account the memory effect arising from the unsteady viscous diffusion of vorticity in the boundary layer surrounding the particle, as also discussed by \citet{Auton1988241} and \citet{Tio1993671}. Depending on the specific flow configuration, some contributions have been neglected with respect to others. For instance, \citet{Coimbra2002257} and \citet{LeProvost2020} neglected selected contributions in their studies of a rotating cylinder and viscous streaming flows, respectively.
\citet{Haller2008573}, by means of a singular perturbation approach, reduced the full equation of motion to a slow manifold capturing the asymptotic dynamics of inertial particles. The resulting reduced-order equation provides an explicit expression for the particle velocity field and has proven useful in practical configurations. \citet{Oettinger2018} employed this formulation to investigate particle dynamics in a T-junction geometry, considering non-interacting gas bubbles, hollow glass beads and polystyrene beads.
Regarding the Basset–Boussinesq history force, this term has been often neglected due to the analytical and computational difficulties of its treatment. However, it becomes significant for finite particle–fluid density ratios in the presence of vortical and oscillatory flows, as recently reported by \citet{Kapoor2024} who performed numerical simulations to examine the trapping of inertial particles in a prototypical two-dimensional rotating flow described by a rotating pair of Lamb–Oseen vortices of identical strength. The BB memory force depends on the entire past evolution of the particle–fluid relative velocity and is classically expressed as a temporal convolution with a weakly singular kernel proportional to $t^{-1/2}$ (among others \citet{Daitche2015567}, who studied the role of the history force on the dynamics of inertial particles in a turbulent flow). As a consequence, inclusion of the BB term transforms the original dynamical system in the position–velocity phase space into an integro-differential equation with non-local dynamics in time, i.e. the entire past trajectory is required to determine a future state of the particle. \citet{Tatom1988283} has also shown that the integral in the BB term can be identified as a fractional derivative of order 1/2, and therefore the motion equation can be considered as a fractional ordinary differential equation. \citet{Prasath2019428} and \citet{Jaganathan2025135} have developed a numerical procedure to efficiently compute the kernel of the relevant integro-differential equation, as outlined by \citet{Haller20191}. More recently, \citet{Legendre2019} and \citet{Gode2023} have derived more complex expressions for the history force, and have shown that for the motion of a drop in an oscillatory flow the classical kernel (proportional to $t^{-1/2}$) is retrieved for very small particle-fluid viscosity ratio, i.e. for the solid particle case.

\subsection{Aim of the work}\label{subsec:aim}

Despite several works in the literature, the full understanding of particle-laden flows in a square microfluidic T-junction still requires further investigations for various flow and geometrical conditions. The present work aims to clarify the dynamics of inertial particles in the T-junction flow configuration by explicitly accounting for the Basset--Boussinesq history force, which has been neglected in most previous studies. The particle force balance equation is used to assess the relative importance of the forces acting on the particles and to determine how particle properties influence accumulation and trapping over a range of flow regimes.

The analysis combines single-phase numerical simulations of the unperturbed base flow for Reynolds numbers in the range $[300,570]$ with integration of the particle force balance equation. For each $Re$, particle trajectories are investigated as functions of the particle-to-fluid density ratio $\rho$ and the Stokes number $St$, defined as the ratio of the viscous diffusion time to the relevant convective time. The effect of the particle injection location over the inlet cross-section is also examined, allowing the fraction of trapped particles to be determined in the $(\rho,St)$ parameter plane and the corresponding basin of attraction to be characterized as well. It will be shown that including the BB contribution reduces the basin of attraction (BoA), indicating that memory effects promote the particle displacement away from the vortex core and weaken the trapping over the whole parameter range  due to the wider spiral trajectories in the radial direction and greater elongation in the axial direction. As will be discussed later, this finding is opposite to that of \citet{Kapoor2024} due to the three dimensional feature of the present configuration.

The investigation is supported by a modal analysis based on spectral proper orthogonal decomposition (SPOD), which is used to identify the unsteady dominant vortical structures at $Re=570$ (i.e., after the Hopf bifurcation) and relate their oscillatory behaviour to particle transport and trapping. In particular, SPOD provides a frequency-based interpretation of the flow unsteadiness and helps establish a direct connection between the breathing dynamics of the primary vortex pair and the phase-dependent particle response observed at $Re=570$, where both the shape and the extent of the basin of attraction depend on the injection time. This study is finally corroborated by a statistical analysis to evaluate the relative importance of the various forces along the trajectory. 

\section{Flow configuration and methodology}\label{sec:methodology}

The geometry under study is that of \citet{Vigolo2013236} and \citet{Chen2017257}, which consists of a $90^\circ$ T-junction channel with square cross-sections, shown in figure~\ref{fig:sketch}. Details on the numerical simulations are given in \S~\ref{subsec:num_modelling}, followed by a description of the equations governing the particle trajectories (\S~\ref{subsec:particles_trajectory}). 

\subsection{Numerical modelling}\label{subsec:num_modelling}

\begin{figure}%
	\centering
\includegraphics[scale=1]{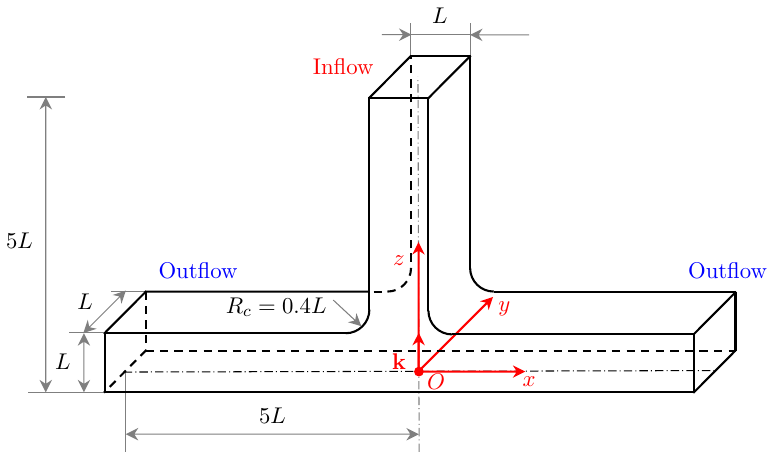}
	\caption{Sketch of the T-junction numerical domain.}
	\label{fig:sketch}
\end{figure}

The analysis is based on direct numerical simulations (DNS) of the incompressible Navier-Stokes equations using the OpenFOAM solver \citep{Weller1998}, which is an open-source fluid dynamics toolbox. The numerical simulations are carried out for a single-phase (base flow), while a force balance equation is used to study the particle motion within the base flow field, assuming that the base flow is unaffected by the particle dynamics. The geometry is fully 3D and discretized using regular hexahedral elements. Note that we consider the entire geometry of the junction (see figure~\ref{fig:sketch}). 

The governing equations are made non-dimensional considering as reference quantities the pipe lateral length $L$, the mean velocity of the flow at the inlet $U$, the dynamic pressure $\rho_f U^2$ ($\rho_f$ being the fluid density) and the corresponding reference time $t_{ref}=L/U$. Indicating the dimensional quantities with the superscript $\left(\cdot\right)^*$, the main dimensionless quantities are:
\begin{equation*}
\mathbf{x}=\mathbf{x^*}/L; \quad \quad \mathbf{u}=\mathbf{u^*}/U; \quad \quad t=t^* U/L; \quad \quad p=p^*/(\rho_f U^2);  \quad \quad \mathbf{g}=\mathbf{k}\mbox{g}^*L/U^2=\mathbf{k}/Fr; 
\label{adimens}
\end{equation*}
where g, $\mathbf{k}$ and $Fr$ are, respectively, the magnitude of the gravitational acceleration, the vertical unit vector and the Froude number $Fr=U^2/g^*L$.

Referring to figure~\ref{fig:sketch}, the inlet and the outlet have the same square cross-section, and the incoming flow is aligned with the vertical direction. The origin of the coordinate system is at the centre of the junction on the bottom plane and the $x$ axis divides symmetrically the horizontal channel and is aligned with the streamwise direction. The $y$ axis extends in the spanwise direction, whereas the $z$ axis points towards the inlet in the vertical direction. Note that $y=0$ coincides with the $x-z$ symmetry plane. 

For the simulations the fluid motion is assumed to be steady and undisturbed. A pressure-driven, laminar, fully developed velocity profile has been imposed at the inlet, defined as an infinite sum (truncated to 200 terms) according to \citet{white2006viscous}. On the solid walls no-slip conditions are imposed, whereas at the outlet sections the relative pressure is set equal to zero $p=0$, with $p$ being the difference between the internal pressure and the ambient reference.
 
\subsection{Particle trajectory}\label{subsec:particles_trajectory}

Let $\mathbf{X^*}(t^*)$ denote the path of a finite-size particle immersed in a constant density fluid moving at velocity $\mathbf{u^*}(\mathbf{x^*}, t^*)$. The motion of a spherical particle of radius $a^*$ and density $\rho_p$ is modelled by the momentum balance equation \citep{Maxey1983883, Gatignol, Tio1993671, Haller2008573}:
  \begin{equation}    	
    \mathbf{v^*}= \frac{d  \mathbf{X^*}}{d t^*},
   \label{eqdxdtd}
\end{equation}

\begin{equation}  
\begin{aligned}
   {\cal V}_p \left(\rho_p+\frac{1}{2}\rho_f\right)\frac{d  \mathbf{v^*}}{d t^*} = & \frac{3}{2}{\cal V}_p \rho_f \frac{D \mathbf{u^*}}{D t^*}+ {\cal V}_p\left(\rho_p-\rho_f\right)\mathbf{g^*}-6 \pi a^* \nu \rho_f (\mathbf{v^*}-\mathbf{u^*})
  \\ & - 6 \sqrt{\pi \nu} (a^*)^2 \rho_f \int_{t^*_0}^{t^*} \frac{1}{\sqrt{t^*-\tau}} \left( \frac{d \mathbf{v^*}}{d\tau} - \frac{d \mathbf{u^*}}{d\tau} \right)  d\tau
   \label{eqpm1d}
   \end{aligned}
 \end{equation}
where $\mathbf{v^*}  = \mathbf{v^*}(\mathbf{x^*}, t^*)$ is the velocity of the particle of volume ${\cal V}_p=\frac{4}{3}\pi {a^*}^3$, $\nu$ is the kinematic viscosity of the fluid, and $t^*_0$ the particles injection time.
The terms on the right-hand side of equation (\ref{eqpm1d}) are, respectively: the force acting on the particle due to the acceleration of the fluid field, the buoyancy force, the viscous drag due to the relative motion between the fluid and the particle, and the Basset-Boussinesq history term. Note that the history force is written in its classical form, namely with the kernel scaling in time as $t^{-1/2}$. The coefficients $\left(\rho_p+\frac{1}{2}\rho_f\right)$ and 3/2 on the left- and the right-hand sides of equation \eqref{eqpm1d} account for, respectively, the inertial added masses of the spherical particle and of the fluid volume.

The inertial force of the fluid, $\frac {D \mathbf{u^*}}{D t^*}$, is given by the fluid flow momentum equation
  \begin{equation}    
   \frac{D  \mathbf{u^*}}{D t^*}=-\frac{1}{ \rho_f} \nabla p^*+\nu \nabla^2 \mathbf{u^*} +\mathbf{g^*}, \label{eqfm1}
\end{equation}
where $D(\cdot)^*/D t^*= (\partial/\partial t^*+ \mathbf{u^*} \cdot \nabla)(\cdot)^*$ is the material derivative.

Let $\rho= {\rho_p}/{\rho_f}$ be the density ratio; then, equation (\ref{eqpm1d}) is recast in the Lagrangian form

\begin{equation}
\label{eqpm2d}
\begin{aligned}
\frac{d \mathbf{v}^*}{d t^*} = & \frac{3}{2} R \frac{D \mathbf{u}^*}{D t^*} + R(\rho - 1)\mathbf{g}^* - \frac{9}{2}\frac{\nu R}{{a^*}^2}(\mathbf{v}^* - \mathbf{u}^*)
  \\ & - \frac{9}{2}\sqrt{\frac{\nu}{\pi}}\frac{R}{a^*} \int_{t^*_0}^{t^*} \frac{1}{\sqrt{t^*-\tau}} \left( \frac{d \mathbf{v}^*}{d\tau} - \frac{d \mathbf{u}^*}{d\tau} \right)\, d\tau,
	   \end{aligned}
\end{equation}
with $R=\dfrac{2}{1+2\rho}$.

In dimensionless form, the equations are    
  \begin{equation}    
   \frac{D  \mathbf{u}}{D t}=- \nabla p+\frac{1}{Re} \nabla^2 \mathbf{u} +\frac{\mathbf{k}}{Fr},  \label{adeqfm1}
\end{equation}

   \begin{equation}    
    \frac{d  \mathbf{v}}{d t}= \frac{3}{2}R \frac{D \mathbf{u}}{D t}+R(\rho-1)\frac{\mathbf{k}}{Fr} -\frac{R}{St}(\mathbf{v}-\mathbf{u}) - \frac{3}{\sqrt{2 \pi}} \frac{R}{\sqrt{St}} \int_{t_0}^{t} \frac{1}{\sqrt{t-\tau}} \left( \frac{d \mathbf{v}}{d\tau} - \frac{d \mathbf{u}}{d\tau} \right)  d\tau    ,  \label{eqpm1}
 \end{equation}
where $a=a^*/L$, $\mathbf{v}=\mathbf{v^*}/U$, and $St$ is the Stokes number here defined as $St=\frac{2}{9} a^2 Re$.

Combining equations (\ref{adeqfm1}) and (\ref{eqpm1}) yields
   \begin{equation} \label{eqpm2}
   \begin{aligned}
    \frac{d  \mathbf{v}}{d t}= & \frac{3}{2}R \left(- \nabla p+\frac{1}{Re} \nabla^2  \mathbf{u} +\frac{\mathbf{k}}{Fr} \right)+R(\rho-1)\frac{\mathbf{k}}{Fr}- \frac{R}{St}(\mathbf{v}-\mathbf{u}) 
    \\ & - \frac{3}{\sqrt{2 \pi}} \frac{R}{\sqrt{St}} \int_{t_0}^{t} \frac{1}{\sqrt{t-\tau}} \left( \frac{d \mathbf{v}}{d\tau} - \frac{d \mathbf{u}}{d\tau} \right)  d\tau.
    \end{aligned}
 \end{equation}

To determine the velocity of particles along their trajectories, as well as the trajectories, we have integrated equation \eqref{eqpm2} in its full form together with the dimensionless equation \eqref{eqdxdtd}. Since equation \eqref{eqpm2} is an integro-differential equation with a singular kernel, the integration has been carried out by means of the explicit numerical procedure recently proposed by \citet{Jaganathan2025135} in MATLAB environment. 

To gain physical insight into the role of the different forces acting on the particles, equation \eqref{eqpm2} has also been studied analytically. By collecting on the left-hand side all the terms involving the particle velocity, the equation can be rewritten as
\begin{equation}
\frac{d\mathbf{v}}{dt} +\frac{R}{St}\mathbf{v} +\frac{3}{\sqrt{2\pi}}\frac{R}{\sqrt{St}} \int_{t_0}^{t}\frac{1}{\sqrt{t-\tau}} \frac{d\mathbf{v}}{d\tau}\,d\tau = \mathbf{F}(t), 
\label{eq:general_BB}
\end{equation}
where
\begin{equation}
\mathbf{F}(t)= \frac{3}{2}R\frac{D \mathbf{u}}{D t}+R(\rho-1)\frac{\mathbf{k}}{Fr}+\frac{R}{St}\mathbf{u}+\frac{3}{\sqrt{2\pi}}\frac{R}{\sqrt{St}}\int_{t_0}^{t}\frac{1}{\sqrt{t-\tau}}\frac{d\mathbf{u}}{d\tau}\,d\tau.
\label{eq:forcing_BB}
\end{equation}
Transforming the convolution integral on the left hand side of equation \eqref{eq:general_BB} into the fractional derivative of the particle velocity, namely introducing the semiderivative  ${d^{1/2}\mathbf{v}}/{d t^{1/2}}$ (  \citet{Tatom1988283})         
\begin{equation}
\int_{t_0}^{t}\frac{1}{\sqrt{t-\tau}}\frac{d\mathbf{v}}{d\tau}\,d\tau = 
\sqrt{\pi}\left[\frac{\mathbf{v}_0}{\sqrt{\pi t}} +\frac{d^{1/2}\mathbf{v}}{d t^{1/2}}   \right],
\label{eq:fra_deriv}
\end{equation}
the solution for the particle velocity can be formally expressed as
\begin{equation}
\mathbf{v}(t)=H(t)\mathbf{v}_0+\int_{t_0}^{t}G(t-\tau)\,\mathbf{F}(\tau)\,d\tau,
\label{eq:formal_solution_BB}
\end{equation}
where \(H(t)\mathbf{v}_0\) is the solution of the corresponding homogeneous problem (i.e. in the absence of forcing) depending on the initial condition $\mathbf{v}_0$, and \(G(t)\) is a kernel associated with the linear operator on the left-hand side of \eqref{eq:general_BB}. Note that \(H(t)\) and \(G(t)\) are not elementary functions, and are generally expressed in terms of Mittag--Leffler functions \citep{Gorenﬂo20201}. Details of the relevant analytical developments are reported in Appendix \S\ref{sec:Mittag}.

This representation shows more clearly that the BB term makes the particle response non-local in time. In fact, when the BB term is ignored, the coefficient \(R/St\) measures the strength of the local drag, so that increasing \(R/St\) promotes a rapid exponential relaxation of \(\mathbf{v}\) to the asymptotic value. However, when the memory effects are included, the particle behaviour is not governed by \(R/St\) alone, but it also depends on the history contribution whose prefactor scales as \(R/\sqrt{St}\). The particle dynamics is therefore controlled by the combined action of inertia, local drag and memory effects, and the transient adjustment of the fluid velocity is not exponential even in cases of large $R/St$, thus producing a longer transient phase. An analytical study of the fractional differential equation, carried out for typical values of $R$ and $St$ (briefly discussed in Appendix \S\ref{sec:Mittag}), shows that the BB history force plays a significant role in most situations, as will be further discussed in Section (\S~\ref{sec:Results}) affecting both the trajectory patterns and the basin of attraction of the injected particles.

On the contrary, if one neglects BB term, as done in most previous works, equation \eqref{eq:general_BB} reduces to the first-order linear ordinary differential equation
\begin{equation}
\frac{d\mathbf{v}}{dt}+\frac{R}{St}\mathbf{v}=\mathbf{\tilde{F}}(t),
\label{eqpm2b}
\end{equation}
where \(\mathbf{\tilde{F}}(t)\) is given by \eqref{eq:forcing_BB} without the history contribution. Analytical integration of \eqref{eqpm2b} gives
\begin{equation}
\mathbf{v}(t)=e^{-\frac{R}{St}(t-t_0)}\mathbf{v}_0
+\int_{t_0}^{t} e^{-\frac{R}{St}(t-\tau)}\,\mathbf{\tilde{F}}(\tau)\,d\tau,
\label{eq:int}
\end{equation}
where \(\mathbf{v}_0\) is the particle velocity, taken equal to the local fluid velocity at the injection time \(t_0\).
Equation \eqref{eq:int}, to be compared with the more general equation \eqref{eq:formal_solution_BB}, shows that the solution is the sum of a transient contribution, associated with the homogeneous equation and exponentially decaying at rate \(R/St\), and a forced contribution, corresponding to a particular solution. Moreover, for \(t-t_0 \gg St/R\), the transient term is strongly damped, so that the particle trajectory is mainly governed by fluid flow inertia (pressure field) and local drag, together with buoyancy effects.
Equation \eqref{eqpm2b}, together with the dimensionless form of \eqref{eqdxdtd}, is integrated in a MATLAB environment by means of a second-order Runge--Kutta method suited to stiff equations.

A further approximate analytical solution of the particle velocity can be obtained by considering the asymptotic solution of equation \eqref{eq:int}. Exploiting the identity ${3R}/{2} = 1-\left(\rho-1\right)R$, one obtains
\begin{equation}
\mathbf{v}=\mathbf{u}+St\left(1-\rho\right)\left(\frac{D \mathbf {u}}{Dt}- \frac{\mathbf{k}}{Fr}\right),
   \label{oett}
	\end{equation}
that reduces to the approximate solution given by \cite{Oettinger2018} when the buoyancy force is neglected.

\section{Results}\label{sec:Results}

For the geometrical configuration under study, we investigate how flow topology, particle size, and density ratio influence particle trajectories, the extent of the particle trapping region, and the associated accumulation and dispersion processes.
A sketch of the geometry is depicted in figure \ref{fig:sketch}. The vertical channel extends $5L$ from the inlet to the bottom plane of the junction. The horizontal branches also extend $5L$ from the origin of the reference frame to the outlet sections. The sides of the junction are rounded with a radius of curvature equal to $R_c=0.4 L$. In the present study we take  $L=4.8\cdot10^{-3}$m corresponding to the experiments of \citet{Vigolo2013236}.

The analysis is carried out for various Reynolds numbers in the range $[300, 570]$. However, we will only discuss the cases $Re=500$, for which vortex breakdown occurs \citep{Vigolo20144770}, and $Re=570$, for which a Hopf bifurcation is observed \citep{Chen2017257}. For each $Re$ the influence of the density ratio, the Stokes number as well as the particle injection location and injection time on the particle dynamic is analysed. To minimize the effects of the numerical initialization procedure, data is analysed starting from $t_0=40$.

\subsection{Vortex flow and particle dynamics}\label{subsec:Vortex_dynamics}

\subsubsection{Case $Re=500$}\label{subsubsec:Re_500}

\begin{figure}%
	\centering
\includegraphics[scale=1]{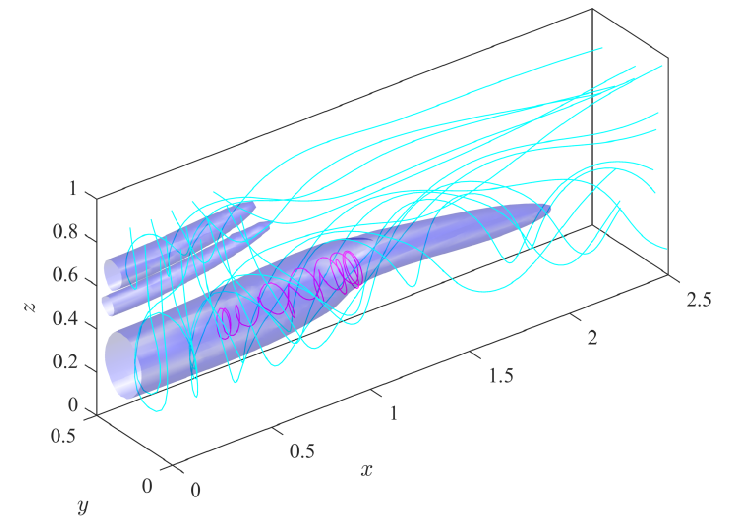}
	\caption{Mean flow at $Re=500$. The flow field is limited to a quarter of the domain. In cyan the streamlines starting from various locations in the inlet channel; the blue surface is the Q-criterion isosurface $Q=25$. Depicted in magenta is a streamline in the inner core of the primary vortex structure.}
	\label{fig:mean_flow_500}
\end{figure}

At $Re=500$ (and $Fr=0.18$ chosen to match the experiments of \citet{Vigolo20144770}) the base flow is dominated by the occurrence of a vortex pair that extends symmetrically with respect to the $x=0$ and $y=0$ planes in the horizontal branches. This test case features spatial complexity of the flow and it has been selected to analyse the dependency of the particle dynamics on the injection location. Figure~\ref{fig:mean_flow_500} reports a Q-criterion isosurface \citep{JEong199569} and various streamlines for $z \le 1$. Note that we report a quarter of the domain only. Once the flow is injected in the vertical branch it moves downstream until the T-bifurcation, then the bottom wall of the junction causes curvature in the streamlines. Under the action of the centrifugal force and the radial pressure gradient, the flow exhibits a spiral motion around the vortex axis moving towards the outlet. Two secondary vortex pairs form in the upper part of the domain, as already shown in the work of \citet{Chen2015}, and close to the curved surface at the branching, where a recirculation zone arise. However, their effects are limited to a small region of the junction. 

The major feature of the flow field is the presence of a bubble-type vortex breakdown \citep{Vigolo20144770, Chen2015}. The latter authors found that for $Re>315$, the breakdown produces four recirculation regions (one for each outlet-depth quadrant) with internal stagnation points on the axes of the vortices. Increasing the Reynolds number, the onset of these vortex structures moves towards the symmetry plane until it is flush with the centre plane ($x=0$). In agreement with these observations, at $Re=500$ the flow is found to exhibit four outlet stagnation points in the horizontal branches, delimiting the vortex breakdown region, and two further stagnation points at the centre of the vortices. The inner streamlines (namely, within the primary pair, reported in magenta in figure \ref{fig:mean_flow_500}) spiral around the vortex axes reversing their direction just upstream of the stagnation point.
 
\subsubsection*{Basin of attraction and particle trajectories}\label{subsubsec:forces}

Equation \eqref{eqpm2} shows that, for given values of $Re$ and $Fr$, particle trajectories depend on the density ratio and the Stokes number. The injection location also affects the individual trajectories, some of which may be trapped in the flow. This enables the identification of a basin of attraction (BoA) in the $(x,y)$ plane defined as the region where particles, once injected, remain trapped and asymptotically approach the axis of the primary vortex on the symmetry plane, which acts as an attractor. 
Before examining individual particle trajectories in detail, we first assess the effect of the BB term on the extension of the basin of attraction injecting particles at $z = 1$ over a range of $(x,y)$ location and solving, respectively, equations~\eqref{eqpm2} and \eqref{eqpm2b}. 

\begin{figure}
	\centering
	\includegraphics[scale=.9]{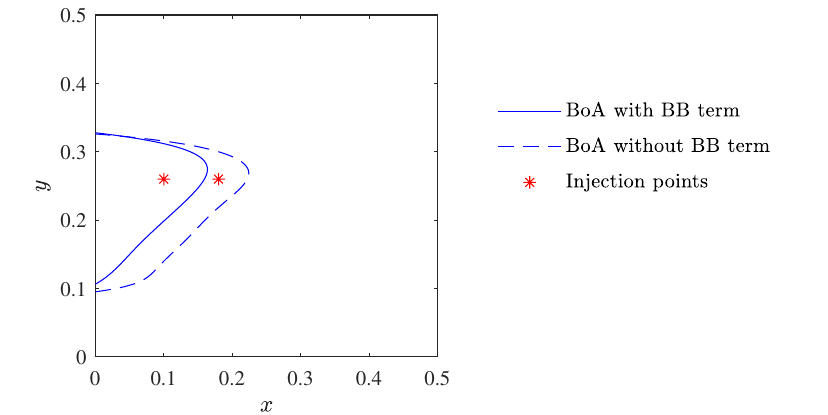}
	\caption{Effect of the Basset–Boussinesq (BB) history term on the extension of the basin of attraction. The figure refers to one quarter of the domain only. Parameters: $Re = 500$, $Fr = 0.18$, $\rho = 10^{-3}$ and $St = 10^{-2}$. Markers denote selected injection points at $x = 0.1$ and $0.18$, with $y = 0.26$.}
	\label{fig:BOA_extension}
\end{figure}

For a given value of $St$ and $\rho$ ($St = 10^{-2}$, $\rho = 0.2$) the region delimited by the solid (respectively, dashed) line in figure  figure~\ref{fig:BOA_extension} identifies the BoA in the presence (respectively, absence) of BB effects.
A relatively sharp boundary is in general observed around $y \approx 0.32$; this can be attributed to the presence of vortical structures close to the wall (as shown in figure \ref{fig:mean_flow_500}) that extend towards the horizontal outlet and advect particles away from the primary vortex, thereby preventing their capture. The figure shows that the effect of the history term is to reduce the size of the BoA. This behaviour is in contrast with the finding of \citet{Kapoor2024} for a prototypal two dimensional rotating pair of Lamb-Oseen vortices, thus highlighting the importance of three-dimensional effects of the present configuration, characterized by strongly complex flow field.

\begin{figure}
	\centering
  \subfloat[]{\includegraphics[scale=.9]{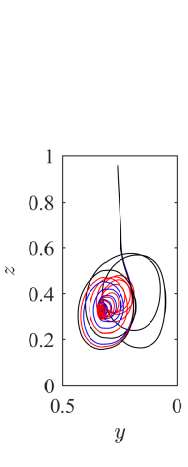}} \hspace{.2cm}
	\subfloat[]{\includegraphics[scale=.9]{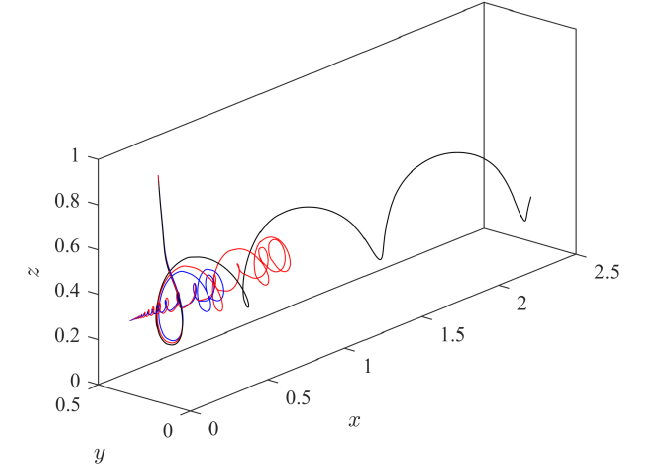}}
\caption{Front (a) and lateral (b) views of the particle trajectories for $\rho = 0.2$ and $St = 10^{-2}$ at $x = 0.1$, $y = 0.26$ and $z = 1$. Red line refers to equation ~\eqref{eqpm2}; blue curve, equation~\eqref{eqpm2b}. In black the streamline starting at the same location. $Re=500$ and $Fr=0.18$.}
	\label{fig:streamlines_St_1e_2_rho_02_tr1}
\end{figure}

\begin{figure}
	\centering
  \subfloat[]{\includegraphics[scale=.9]{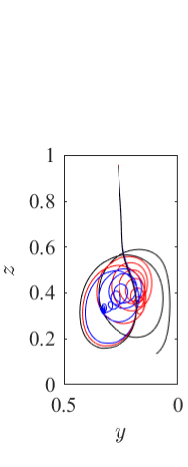}} \hspace{.2cm}
	\subfloat[]{\includegraphics[scale=.9]{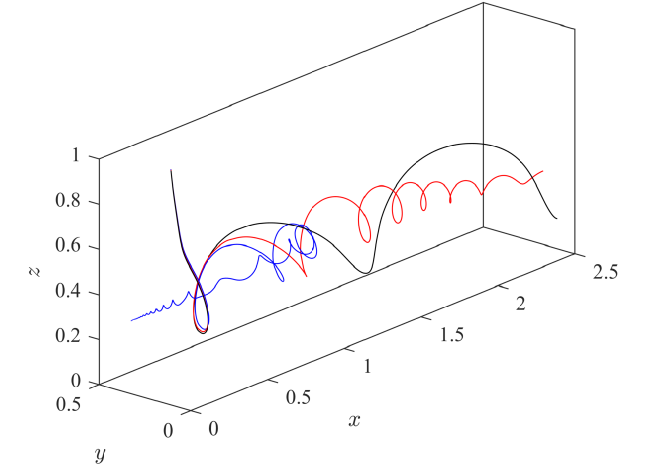}}
\caption{Front (a) and lateral (b) views of the particle trajectories for $\rho = 0.2$ and $St = 10^{-2}$ at $x = 0.18$, $y = 0.26$ and $z = 1$. Red line refers to equation ~\eqref{eqpm2}; blue curve, equation~\eqref{eqpm2b}. In black the streamline starting at the same location. $Re=500$ and $Fr=0.18$.}
	\label{fig:streamlines_St_1e_2_rho_02_tr2}
\end{figure}

To further assess the influence of the history force, particle trajectories computed starting from the same injection point, with and without the BB term, are compared. Figures~\ref{fig:streamlines_St_1e_2_rho_02_tr1} and \ref{fig:streamlines_St_1e_2_rho_02_tr2} depict trajectories from injection points located, respectively, at $x = 0.1$ and $0.18$, with $y = 0.26$ and $z = 1$, corresponding to the markers in figure~\ref{fig:BOA_extension}. The red line represents the trajectory  computed by solving the equation~\eqref{eqpm2}, the blue curve is the trajectory corresponding to equation~\eqref{eqpm2b}, while the black line denotes the flow streamline, whose origin is at the injection location. It is interesting to observe that the BB force acts as an additional drag, leading the trajectories not only to have a larger radius of curvature (figure~\ref{fig:streamlines_St_1e_2_rho_02_tr1}$a$), but also a greater downstream extension (figure~\ref{fig:streamlines_St_1e_2_rho_02_tr1}$b$). As a result, particles tend to remain farther from the vortex core, thereby reducing the likelihood of being trapped and, consequently, the extent of the BoA. This behaviour is further illustrated in figure~\ref{fig:streamlines_St_1e_2_rho_02_tr2}, which shows that the BB force induces a sufficiently strong outward displacement for the particle to escape the vortical region, whereas in the absence of the memory term the particle remains trapped.

\newpage
\subsubsection*{Particle balance equation in a cylindrical reference system}\label{subsubsec:cylindr_frame}

\begin{figure}%
	\centering
\includegraphics[scale=0.95]{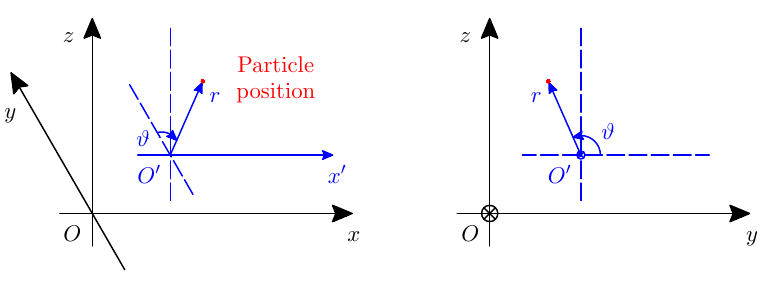}
	\caption{Sketch of the local cylindrical reference system. a) global view; b) view in the $y-z$ plane.}
	\label{fig:sketch_rs}
\end{figure}

The basic physical mechanisms responsible for particle trapping are well established (among others, \citet{Vigolo20144770}). They depend on the interplay between pressure gradient, centrifugal force and viscous drag in the radial direction and pressure gradient and  viscous drag along the axial one. In the present work, to elucidate the role played by the history force, we reformulate the balance equation \eqref{eqpm2} in a cylindrical reference system centred at the vortex core and analyse the trend of the various forces acting on the particles. For that purpose, at a time instant $t$, once the vortex core is identified (by searching for the minimum of pressure in a $y-z$ plane containing the particle), we construct a local cylindrical reference system with the longitudinal axis $x'$ parallel to $x$ and the radial ($r$) and azimuthal ($\vartheta$) axes in the $y-z$ plane. The pair ($r$, $\vartheta$) represents the particle distance from the vortex core and its angular position with respect to the $y$ axis, respectively. A sketch of the local cylindrical reference system is depicted in figure~\ref{fig:sketch_rs}. Observe that the cylindrical frame is not used to integrate the particle equations of motion. Particle trajectories are computed in the original frame, while the cylindrical representation is introduced a posteriori only to project and analyse the different force contributions relative to the vortex core.

Let the subscripts $r$, $\vartheta$, $x$ stand for radial, angular and longitudinal components. The balance equation \eqref{eqpm2} expressed in the ($r$, $\vartheta$, $x'$) system reads
\begin{align}    
\begin{aligned}
\label{eq_r}
\frac{d  v_r}{d t}= \frac{{v_{\vartheta}}^2}{r}& + \frac{3}{2} R \left[ \frac{{u_{\vartheta}}^2}{r}- \frac{\partial p}{\partial r} +\frac{1}{Re} \left(\nabla^2 u_r- \frac{u_r}{r^2} -\frac{2}{r^2}\frac{\partial u_{\vartheta}}{\partial \vartheta} \right)\right]  -\frac{R}{Fr}(\rho-1) \sin{\vartheta}\\&- \frac{9}{2}\frac{R}{Re}\frac{1}{a^2} (v_r-u_r)
- \frac{3}{\sqrt{2 \pi}} \frac{R}{\sqrt{St}} \int_{0}^{t} \frac{1}{\sqrt{t-\tau}} \left( \frac{d \mathbf{v_r}}{d\tau} - \frac{d \mathbf{u_r}}{d\tau} \right)  d\tau. 
  \end{aligned}
  \end{align}
\begin{align}    
\begin{aligned}
\label{eq_theta}
\frac{d  v_{\vartheta}}{d t}= & 2 \frac{{v_r v_{\vartheta}}}{r}+\frac{3}{2} R \left[ - \frac{{u_r u_{\vartheta}}}{r}- \frac{1}{r} \frac{\partial p}{\partial \vartheta} +\frac{1}{Re} \left(\nabla^2 u_{\vartheta}- \frac{u_{\vartheta}}{r} + 2r\frac{\partial u_{\vartheta}}{\partial r} \right)\right] \\ -\frac{R}{Fr}(\rho-1) & \cos{\vartheta} - \frac{9}{2}\frac{R}{Re}\frac{1}{a^2} (v_{\vartheta}-u_{\vartheta}) - \frac{3}{\sqrt{2 \pi}} \frac{R}{\sqrt{St}} \int_{0}^{t} \frac{1}{\sqrt{t-\tau}} \left( \frac{d \mathbf{v_{\vartheta}}}{d\tau} - \frac{d \mathbf{u_{\vartheta}}}{d\tau} \right)  d\tau.  
  \end{aligned}
  \end{align}
	\begin{align}    
\begin{aligned}
\label{eq_x}
\frac{d  v_x}{d t}= &\frac{3}{2} R \left[ - \frac{\partial p}{\partial x} +\frac{1}{Re} \left(\nabla^2 u_x \right)\right] - \frac{9}{2}\frac{R}{Re}\frac{1}{a^2} (v_x-u_x)\\&
- \frac{3}{\sqrt{2 \pi}} \frac{R}{\sqrt{St}} \int_{0}^{t} \frac{1}{\sqrt{t-\tau}} \left( \frac{d \mathbf{v_x}}{d\tau} - \frac{d \mathbf{u_x}}{d\tau} \right)  d\tau. 
  \end{aligned}
  \end{align}

\begin{figure}
\captionsetup[subfigure]{labelformat=empty}
	\centering
  \subfloat[]{\includegraphics[scale=.95]{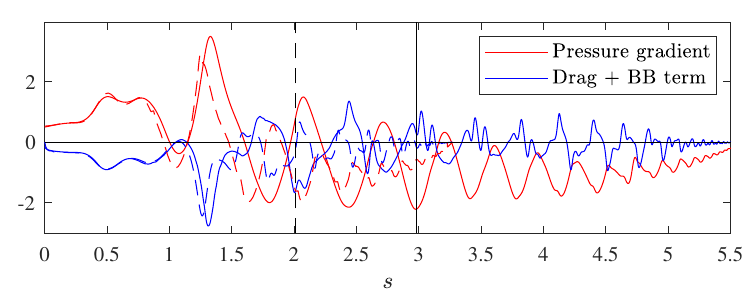}}\vspace{-.7cm}
  \subfloat[]{\includegraphics[scale=.95]{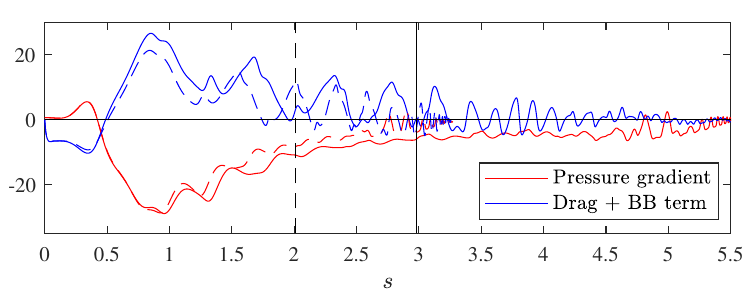}}\vspace{-.7cm} 
  \caption{Force balance terms, \eqref{eq_r} and \eqref{eq_x}, along the axial (top panel) and radial (bottom panel) directions, $\rho=0.2$ and $St = 10^{-2}$. Particle injected at $x = 0.1$, $y = 0.26$ and $z = 1$; $Re=500$ and $Fr=0.18$. Solid lines refer to equation~\eqref{eqpm2}; dashed lines to \eqref{eqpm2b}.} 
	\label{fig:balances_1}
\end{figure}

\begin{figure}
\captionsetup[subfigure]{labelformat=empty}
	\centering
 \subfloat[]{\includegraphics[scale=.95]{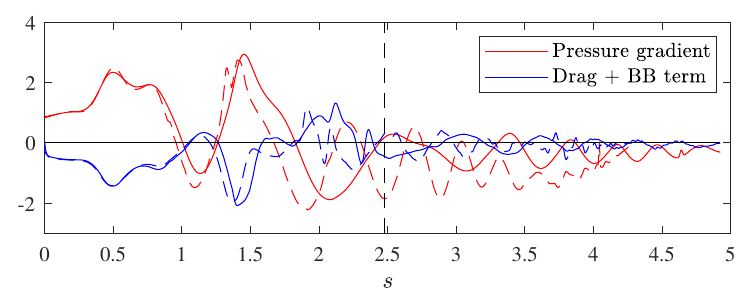}}\vspace{-.7cm}
 \subfloat[]{\includegraphics[scale=.95]{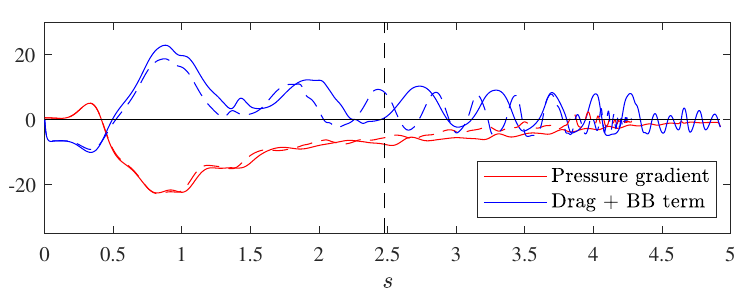}}\vspace{-.7cm} 
			\caption{Force balance terms, eqs. \eqref{eq_r} and \eqref{eq_x}, along the axial (top panel) and radial (bottom panel) directions. $Re=500$ and $Fr=0.18$, $\rho=0.2$ and $St = 10^{-2}$. The injection point is $x = 0.18$, $y = 0.26$ and $z = 1$. Solid lines refer to equation~\eqref{eqpm2}; dashed lines to \eqref{eqpm2b}. Vertical lines indicate the position of motion reversal.}
	\label{fig:balances_2}
\end{figure}

For the analysis, we consider particles injected at the two positions reported in figure~\ref{fig:BOA_extension}. The evolution of the forces acting on the particles along their trajectories is shown in figures~\ref{fig:balances_1} (injection location $x=0.1$) and \ref{fig:balances_2} ($x=0.18$), where $s$ denotes the curvilinear coordinate measured from the injection point. The figures reports the distribution of the pressure gradient and the sum of the local drag and the BB force along the trajectories evaluated with equations~\eqref{eqpm2} (solid lines) and \eqref{eqpm2b} (dashed lines), respectively. The vertical lines indicate the position at which the particle reverses its motion along the streamwise direction.

Considering first the injection point at $x = 0.1$, $y = 0.26$, and $z = 1$, the particle initially moves along a vertical straight path, with drag and BB forces opposing its motion. 
Close to the bottom plane (namely, approximately for $s > 0.5$), the particle begins to describe a spiraling path, which enhances the radial force components, making them dominant in this stage of the trajectory. 
A competition is thus established between radial pressure gradient, local drag and memory force, and centrifugal forces (not shown in these figures for simplicity), which govern the particle dynamics. In particular, for this injection point that corresponds to a trapping case, the centrifugal force initially dominate, pushing the particle away from the vortex core, as is observed in figure~\ref{fig:streamlines_St_1e_2_rho_02_tr1}. However, at later times the pressure gradient prevails, redirecting the particle inward. Once the particle has entered the vortical region, it moves along a spiralling orbit with progressively decreasing curvature radius and finally, due to the prevailing axial pressure gradient, its motion reverses. From this point onwards, the motion is primarily governed by the axial pressure gradient which prevails against drag and BB force (figure~\ref{fig:balances_1}, top panel), and directs the particle towards the symmetry plane $x=0$ along the vortex axis, acting as an attraction region. 

By comparing the force distributions with and without the memory term, it can be observed that, due to its viscous nature, the Basset--Boussinesq term adds to the drag (see blue lines in both panels of figures~\ref{fig:balances_1}). As a result, the overall viscous force increases and acts in opposition to the pressure gradient. This behaviour is observed in both the radial and axial directions. Consequently, the inclusion of the BB term leads to a particle trajectory that is wider in the radial direction and more extended in the axial direction (before the point of motion reversal), as shown by the red curves in figure~\ref{fig:streamlines_St_1e_2_rho_02_tr1}.

Considering now the other injection point under study located at $x = 0.18$, $y = 0.26$ and $z = 1$ (figure \ref{fig:balances_2}) the action of the forces on the particle is essentially the same. In this case, however, the effect of the BB term in extending axially the trajectory displaces the particle far downstream in the recirculation region. As a result, the particle although is pushed inward by the pressure gradient, it can no longer be trapped and is therefore expelled outwards.

\subsubsection*{Stokes number and density ratio effects}\label{subsubsec:St_rho_effects}

\begin{figure}
	\centering
 \subfloat[]{\includegraphics[scale=.9]{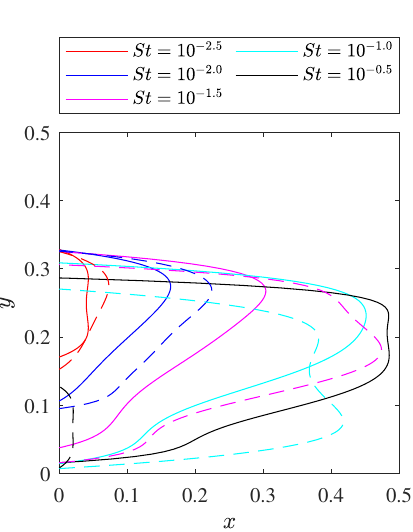}} \hspace{.2cm}
	\subfloat[]{\includegraphics[scale=.9]{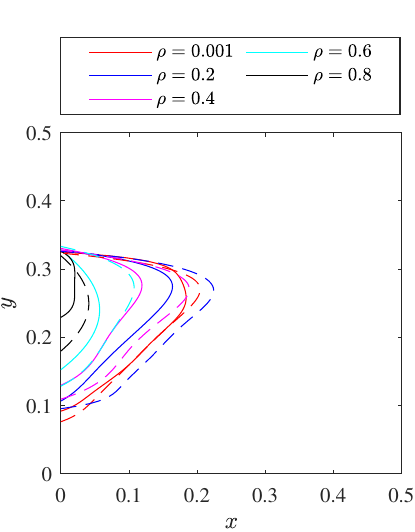}}
\caption{Effect of Stokes number at $\rho=0.2$ (left panel) and density ratio at $St = 10^{-2}$ (right panel) on BoA extent. Solid lines correspond to the full equation~\eqref{eqpm2}, whereas dashed lines correspond to equation~\eqref{eqpm2b} (BB neglected). Only one quarter of the domain is shown. $Re = 500$, $Fr = 0.18$.}
	\label{fig:St_rho_variation}
\end{figure}

The effect of Stokes number and density ratio on the extent of the basin of attraction is reported in figure~\ref{fig:St_rho_variation} for five different values of $St$ and $\rho$, spanning the range from small to large $St$ and from light to heavy particles. Solid lines denote results obtained including the BB force, whereas dashed lines correspond to the ones obtained neglecting the memory term.

    The figure shows that for a fixed $\rho$ value, for small $St$ particles behave nearly as passive tracers, with trajectories closely following the streamlines and spiralling towards the outlet, resulting in a limited BoA extent. Increasing the Stokes number leads to an expansion of the BoA, indicating that particles with larger inertia are more prone to be trapped within the vortical region. On the other hand, increasing the density ratio for a fixed $St$ results in a reduction of the BoA, as heavier particles tend to escape more readily from the vortex core. For $\rho > 1$, particle motion approaches that of freely falling heavy objects, and particles may impact the bottom wall under the action of gravity, thereby preventing trapping. Overall, the inclusion of the BB term leads to a reduction of the BoA, with the corresponding curves being entirely contained within those obtained in the absence of memory effects. This indicates that the latter effects promote particle displacement away from the vortex core, thereby weakening trapping across the entire parameter range. As already observed, this finding reflects the relative importance of inertial effects (essentially the pressure gradient), drag, and memory effects, which jointly govern the particle trapping within the vortical flow.

The effects of the Stokes number can be better understood by noting that $St$ represents the ratio of the viscous diffusion time, 
 $t_v=a^{*2} / \nu$ to the convective time, $t_c=L/U$, namely
\begin{equation}    
      St \approx a^2 Re =\frac{U}{L} \frac{a^{*2}}{\nu} = \frac{t_v}{t_c}.
\end{equation}
When $St$ is small, the time scale of momentum diffusion over regions of the order of $a^*$ is short compared to the time to cover the length $L$, and the particle drag due to fluid-particle relative motion counteracts the pressure gradient directed inwards. In this case, the pressure gradient in the radial direction is not sufficient to balance the centrifugal force acting on the particle preventing its trapping. As the Stokes number increases, the fluid-particle drag acts over a prolonged characteristic time causing the particle trajectories, driven by the prevailing pressure gradient, to move towards the vortex core.

\begin{figure}
	\centering
  \subfloat[]{\includegraphics[scale=.9]{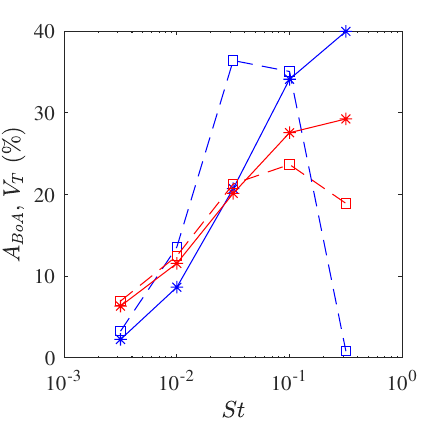}} \hspace{.2cm}
	\subfloat[]{\includegraphics[scale=.9]{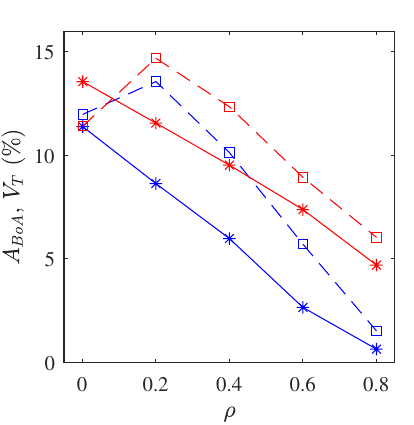}}
\caption{Effect of the Stokes number at at $\rho=0.2$ (left panel) and the density ratio $St = 10^{-2}$ (right panel) on the BoA area ($A_{BoA}$, blue curves) and trapping volume ($V_{T}$, red curves). Solid lines and $*$ markers correspond to equation~\eqref{eqpm2}, whereas dashed lines and $\square$ markers correspond to equation~\eqref{eqpm2b}. Only one quarter of the domain is shown. $Re = 500$, $Fr = 0.18$.}
	\label{fig:St_rho_AVT}
\end{figure}

The effect of the BB term on the particle dynamics has also been examined through the variation of BoA area ($A_{BoA}$) and trapping volume ($V_T$) as a function of $St$ and $\rho$, where the trapping volume is defined as the region of the flow containing all trapped particles trajectories such that a particle injected within the BoA remains trapped in this region. Figure~\ref{fig:St_rho_AVT} shows the variation of $A_{BoA}$ and $V_T$ with the Stokes number at $\rho=0.2$, and with the density ratio at $St=10^{-2}$. Solid lines and $*$ markers refer to results obtained including the BB term, whereas dashed lines and $\square$ markers denote the cases without memory effects.

As the Stokes number increases, the BoA area grows (panel a), with values obtained without the BB term generally larger than those including memory effects. This trend persists up $St \approx 10^{-1}$. Beyond this value, a marked difference emerges: while the BoA area and the trapping volume obtained with the BB term continues to increase, the corresponding values without memory exhibit a pronounced decrease. In contrast, the variation with the density ratio exhibits an opposite trend (panel b). As $\rho$ increases, the number of trapped particles rapidly decreases, leading to a progressive reduction of both $A_{BoA}$ and $V_T$. For sufficiently large density ratios, both quantities tend to vanish, indicating a complete suppression of particle trapping. This behaviour reflects the increasing dominance of particle inertia, since heavier particles are unable to follow the vortical motion required for trapping. As a consequence, they more readily escape the recirculation region, resulting in loss of trapping efficiency with increasing $\rho$. 

\begin{figure}
	\centering
\includegraphics[scale=1]{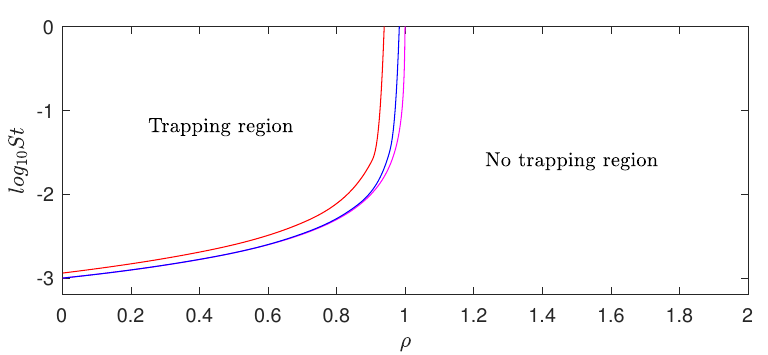}
\caption{Numerical threshold curves obtained solving the equations \eqref{eqpm2} (red curve) and \eqref{eq:int} (blue curve); approximate analytical threshold curve $St(1-\rho)=0.001$ by \citet{Oettinger2018} is represented by magenta line. $Re=500$ and $Fr=0.18$.}
	\label{fig:St_rho}
\end{figure}

\begin{figure}
	\centering
  \subfloat[]{\includegraphics[scale=.9]{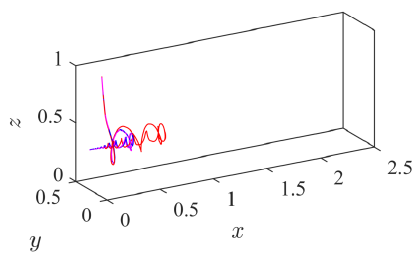}} \hspace{0.5cm}
	\subfloat[]{\includegraphics[scale=.9]{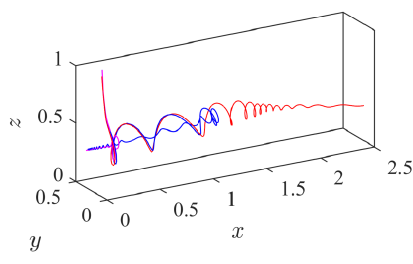}}\\
\caption{Particles trajectories for two values of $R/St$. a) $R/St=200$ ($St=10^{-2}$, $\rho=10^{-3}$); b) $R/St=8.3$ ($St=10^{-1}$, $\rho=0.8$). Red line, numerical solution of equation \eqref{eq:general_BB}; blue line, solution of equation \eqref{eq:int}, magenta line, solution of \eqref{oett}. $Re=500$ and $Fr=0.18$.}
	\label{fig:streamlines_OH}
\end{figure}

An important and practical global objective of the study of the particle dynamics in a carrier fluid is identifying in the ($\rho$, $St$) plane the region where particles are trapped, or, in other words, the threshold curve that separates the region where trapping occurs from the one where particles are not captured.
In figure~\ref{fig:St_rho} we report in the ($\rho$, $St$) plane, the threshold curve obtained by solving the complete equation of motion \eqref{eqpm2} (red line), and the corresponding curve in the absence of memory effects (blue line), obtained by employing equation \eqref{eq:int}. In magenta, we also show the approximate asymptotic prediction by \citet{Oettinger2018}, based on \eqref{oett}. This latter is expressed by the branch of the hyperbola $St(1-\rho)=0.00195$. The threshold curve associated with the solution of the complete equation is entirely contained within that obtained without the memory term, reflecting the tendency of the BB force to promote particle escape. It is interesting observe that since for a fixed $\rho$ the $St$ threshold value is higher, this is equivalent to considering that the inclusion of BB force increases the viscous time, and as a consequence the memory effects along the trajectory slow down the local drag effect.

This behaviour is shown in detail by inspection of the particle trajectories corresponding to \eqref{eq:general_BB}, \eqref{eq:int} and \eqref{oett}, and reported in figure~\ref{fig:streamlines_OH} for \(R/St=200\) (\(St=10^{-2}\), \(\rho=10^{-3}\)) and \(R/St=8.3\) (\(St=10^{-1}\), \(\rho=0.7\)).
The figure shows that the memory term has a significant influence at both high and low values of \(R/St\), whereas the asymptotic solution coincides with that obtained without the BB term for high \(R/St\) (left panel), but departs from it as \(R/St\) decreases (right panel).

\subsubsection{Case $Re=570$}\label{subsubsec:hopf_bifurcation}

This test case has been previously analysed by \citet{Chen2015} who have focused on the instability and sensitivity of pipe flows that undergo a Hopf bifurcation at $Re=556$. In the present paper this test case has been selected for its spatial and temporal complexity to analyse the influence of the injection time on the dynamics of the particles. 

\begin{figure}
\captionsetup[subfigure]{labelformat=empty}
	\centering
  \subfloat[]{\includegraphics[scale=1]{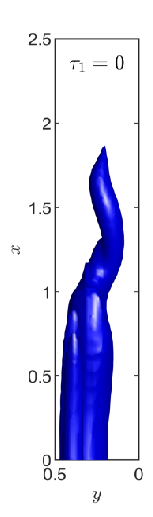}}
  \subfloat[]{\includegraphics[scale=1]{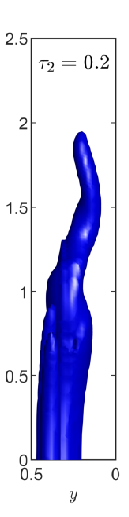}}
  \subfloat[]{\includegraphics[scale=1]{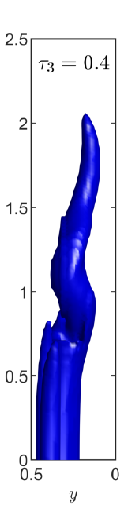}}
  \subfloat[]{\includegraphics[scale=1]{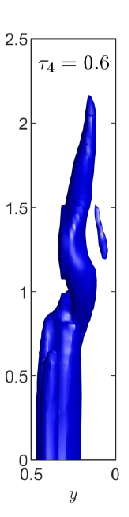}}
  \subfloat[]{\includegraphics[scale=1]{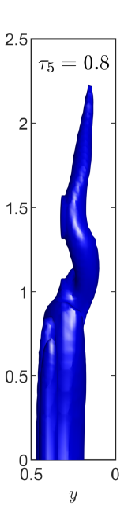}}
	\subfloat[]{\includegraphics[scale=1]{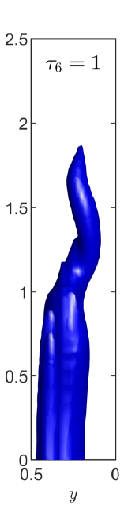}}
  \caption{Time evolution of the flow at $Re=570$ at various dimensionless times $\tau_n$ in the breathing period. The blue surface is a Q-criterion isosurface $Q=25$.}	\label{fig:breathing_top}
\end{figure}

At $Re = 570$ (and $Fr = 0.24$), i.e. after the occurrence of Hopf bifurcation, the base flow exhibits a helical oscillation of the primary vortex, characterised by a corkscrew motion. During this process, the vortex undergoes a periodic breathing, assuming a knee-like shape as it stretches while bending. This behaviour is illustrated in figure~\ref{fig:breathing_top}, where Q-criterion isosurfaces are reported at different time instants within one oscillation period $T$, defined as $t_n / T=t_0/T+ \left( n-1 \right) \Delta t /T $, with $n=1$ $-$ $6$ and $\Delta t/T =0.2$.

\begin{figure}%
	\centering
\includegraphics[scale=.95]{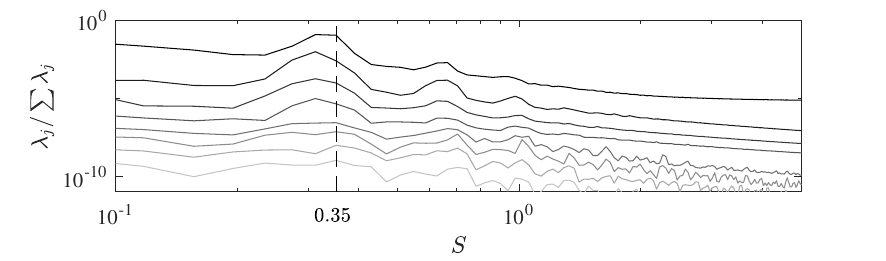}
	\caption{Spectral eigenvalues as a function of the Strouhal number for $Re = 500$.}
	\label{fig:spectrum_500_p}
\end{figure}

\begin{figure}%
	\centering
\includegraphics[scale=.95]{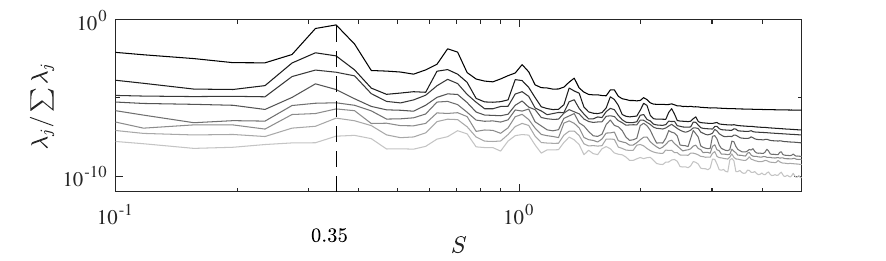}
	\caption{Spectral eigenvalues as a function of the Strouhal number for $Re = 570$.}
	\label{fig:spectrum_570_p}
\end{figure}

To characterise the frequency content of these oscillations, a spectral proper orthogonal decomposition (SPOD) of the pressure field has been performed with the methodology described by \citet{Chiatto2023}, including both Reynolds numbers ($Re = 500$ and $570$) for completeness. The resulting spectra are reported in figures~\ref{fig:spectrum_500_p} and \ref{fig:spectrum_570_p}, respectively, as a function of the Strouhal number ($S=f L/U$, $f$ being the frequency). In both cases, a fundamental frequency, $S = 0.35$, is identified, while, for $Re = 570$, additional super-harmonics are clearly present; furthermore, a clear separation between the leading and higher modes indicates a low-rank behaviour of the system.

\begin{figure}
	\centering
  \subfloat[$Re=500$]{\includegraphics[scale=.9]{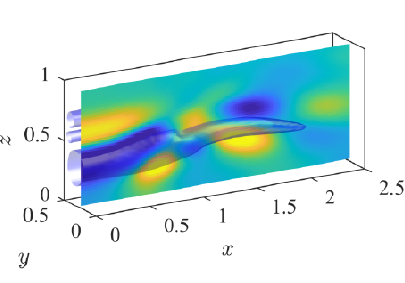}} \hspace{0.1cm}
	\subfloat[$Re=570$]{\includegraphics[scale=.9]{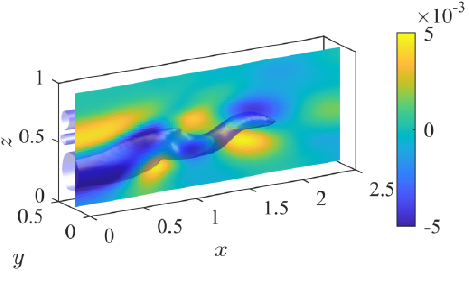}}\\
  \caption{Representation of the real part of the peak Fourier component of the $1^{st}$ SPOD mode at $Re=500$ (left panel) and $Re=570$ (right panel) in the $y=0.5$ plane intersecting the Q-criterion isosurface $Q=25$. The scale of the colormap pressure SPOD mode is also reported.}	
	\label{fig:p_spod_Q_criterium}
\end{figure}

The spatial structures associated with the dominant frequency of the first SPOD mode are shown at $y = 0.25$ in figure~\ref{fig:p_spod_Q_criterium}, together with the $Q$-criterion isosurface ($Q = 25$), for both cases $Re = 500$ (left panel) and $Re = 570$ (right panel). Although the two flow regimes differ significantly, the spatial distribution of the pressure mode is remarkably similar in both cases. In particular, for $Re = 570$, a clear correspondence is observed between the $Q$-isosurface (shown at time $\tau_1$) and regions of local pressure minima, indicating that the peak of the SPOD spectrum is associated with the corkscrew motion of the primary vortex pair, thereby providing information on the characteristic frequency of the breathing dynamics. For $Re = 500$, the flow reaches a steady state and the $Q$-criterion isosurface becomes time-independent. The presence of the same spectral component is attributed to the transient evolution of the associated structures, whose energy content decays slowly. It is therefore expected that, over sufficiently long observation times, this oscillatory contribution vanishes.

To give physical relevance to the peak frequency predicted by SPOD, it is useful to recall that a jet issuing from a slit with mean velocity ($U_0$) and impacting on an edge parallel to the slit produces the so-called edge-tones \citep{Curle1953}, which are associated with the amplification of oscillations due to the interaction of eddies with the edge. The frequencies ($f_{et}$) of these (purely hydrodynamic) tones depend on the eddy's convection velocity ($U_c$) and the jet-tone wavelength ($\lambda$), which is the distance between two consecutive vortices and corresponds to the oscillations of maximum amplitude \citep{Curle1953}, namely
 $f_{et}=U_c/\lambda$, where $\lambda= d/(i+0.25)$ ($d$ being the distance of the edge from the slit and $i$ the frequency stage). 
For the flow under investigation, we can link the shedding of the vortices generated at the end of the recirculation bubble and issuing in a constant ambient pressure at the junction outlet section, to the edge-tones phenomenon. Then, if one assumes a typical value of $U_c = 0.5 U$ and takes for $d$ the value determined from the Q-criterion isosurface at the instant when the vortices are shed (for which we have $d=3.15 L$), one obtains $S_{et}=f_{et} L/U \approx 0.357$ when considering the second stage of Curle's formula (i.e., $i=2$), which is in very good agreement with the peak value predicted by SPOD.

\subsubsection*{Particle trajectories and trapping analysis}\label{subsubsec:forces_570}

\begin{figure}%
	\centering
\includegraphics[scale=.95]{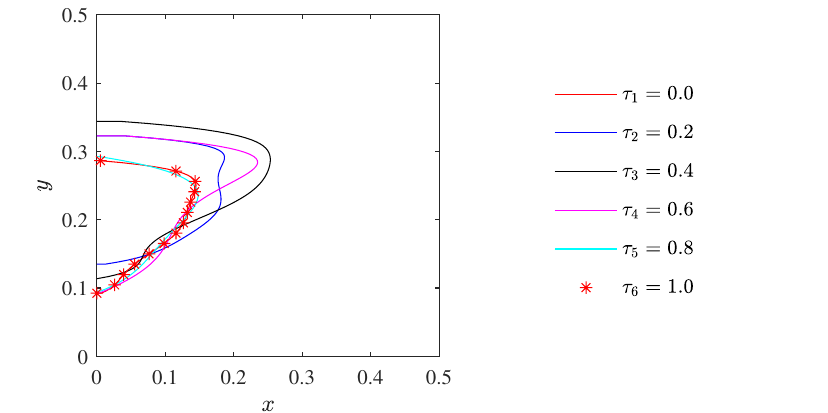}
	\caption{Time evolution of the BoA for $St=10^{-2}$ and $\rho=0.2$ at various dimensionless times $\tau_n$ in the breathing period. $Re=570$.}
	\label{fig:BoA_rho_02_St_1e2_Re_570}
\end{figure}

The periodic breathing motion has a significant impact on particle dynamics and, consequently, on the trajectories of particles injected into the flow field. For a given injection position, the particle may interact with the primary vortical structure at different phases of the breathing cycle depending on the injection time, thereby leading to markedly different trajectories. This introduces a pronounced sensitivity to the initial phase of injection, resulting in a phase-dependent particle dynamics which is also reflected in both the shape and the extent of the BoA, as shown in figure~\ref{fig:BoA_rho_02_St_1e2_Re_570} for $St = 10^{-2}$ and $\rho = 0.2$. Starting from $\tau_1$, the BoA initially expands with increasing the phase, and slightly shifts towards larger $y$ values; it reaches a maximum extension around $\tau_3$, and then gradually returns to its initial shape. It is worth noting that, unlike the $Re = 500$ case, where $y \approx 0.32$ seems a common upper boundary under variations of both $St$ and $\rho$, here the BoA boundaries along $y$ vary with the phase of the breathing cycle. This behaviour can be attributed to the changes in the vortical structures developing near the junction walls, induced by the motion of the primary vortex.

\begin{figure}
\captionsetup[subfigure]{labelformat=empty}
	\centering
  \subfloat[]{\includegraphics[scale=.9]{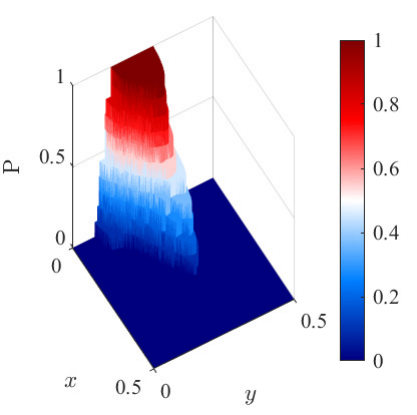}} \hspace{0.1cm}
	\subfloat[]{\includegraphics[scale=.9]{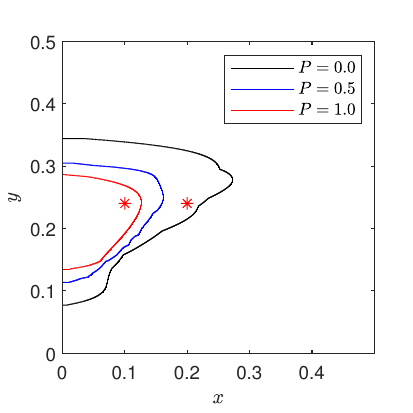}}\\
  \caption{Particle trapping probability over the injection plane (left panel) and the corresponding isolevels for $P = 0$, $0.5$ and $1$ (right panel). $St = 10^{-2}$ and $\rho = 0.2$.}	
	\label{fig:probability}
\end{figure}

The injection phase dependence discussed above can be further quantified in terms of the trapping probability, as shown in figure~\ref{fig:probability}. The left panel reports the probability of particle trapping occurrence over the injection plane ($P$), computed by varying the injection time over the breathing cycle. The right panel shows the corresponding isolevels for $P = 0$, $0.5$ and $1$. The figure highlights the coexistence of regions of certain trapping, certain escape and intermediate probability. In particular, the region for which $0 \leq P \leq 0.5$ identifies the set of injection locations for which the trapping event is most sensitive to the phase of the breathing motion. This behaviour is consistent with the phase-dependent variations of the BoA discussed above, and reflects the periodic displacement of its boundaries during the breathing cycle.

\begin{figure}
\captionsetup[subfigure]{labelformat=empty}
	\centering
  \subfloat[]{\includegraphics[scale=1]{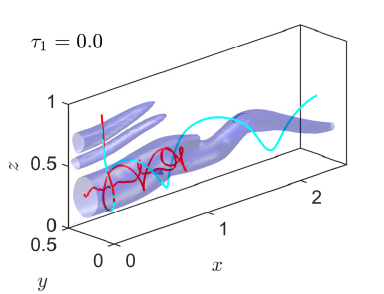}}
  \subfloat[]{\includegraphics[scale=1]{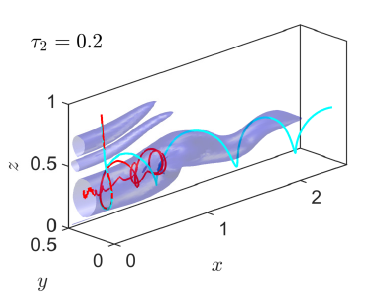}}\vspace{-0.9cm}\\
  \subfloat[]{\includegraphics[scale=1]{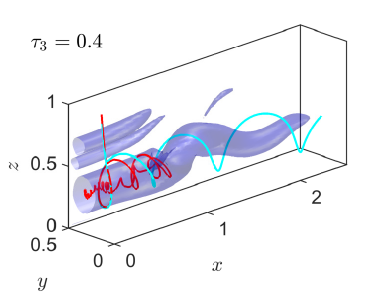}}
  \subfloat[]{\includegraphics[scale=1]{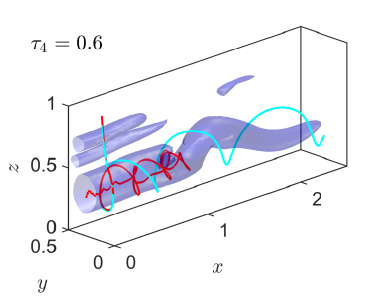}}\vspace{-0.9cm}\\
  \subfloat[]{\includegraphics[scale=1]{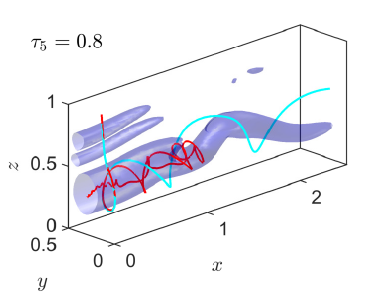}}
	\subfloat[]{\includegraphics[scale=1]{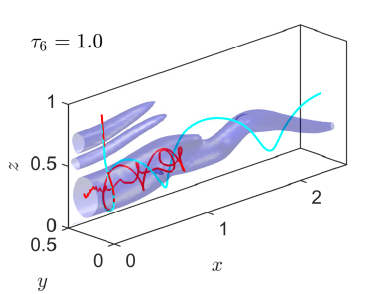}}\vspace{-0.5cm}\\
  \caption{Time evolution of the flow and particle motion at $Re=570$ at various dimensionless times $\tau_n$ in the breathing period. The flow field is limited to a quarter of the domain. Red lines refer to $St=10^{-2}$ and $\rho=0.2$, cyan curves represent the flow pathlines starting at the same location, and the blue surface is a Q-criterion isosurface $Q=25$.}	
  \label{fig:breathing}
\end{figure}

\begin{figure}%
	\centering
\includegraphics[scale=1]{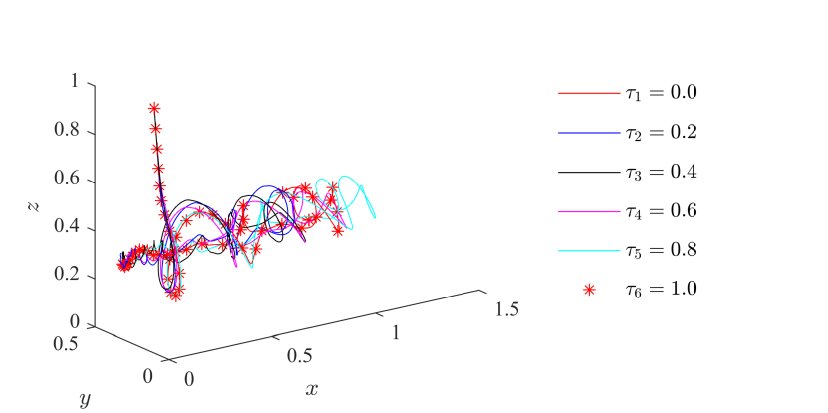}
	\caption{Trajectories of particles injected at $x=0.1$ $y=0.26$ and $z=1$ (with $St=10^{-2}$ and $\rho=0.2$) at various dimensionless times $\tau_n$ in the breathing period. $Re=570$ and $Fr=0.24$.}
	\label{fig:trajectories_570}
\end{figure}

To corroborate the analysis of the effect on the motion of trapped particles of the unsteadiness of base flow produced by the Hopf bifurcation at $Re=570$, we have injected $n$ particles (with $n=1$ $-$ $6$, $St=10^{-2}$ and $\rho=10^{-3}$) at the same spatial location $x = 0.1$, $y = 0.26$, $z = 1$ and at various time instances $\tau_n$ in the breathing period (the instances being those selected for the flow analysis previously discussed). The corresponding trajectories  (red lines) and Q-criterion isosurfaces are reported in figure~\ref{fig:breathing}. For clarity, we also include in figure~\ref{fig:trajectories_570} the inner trajectories at all $t_n$. 

These figures show that the position at which the particles reverse their motion initially moves towards the symmetry axis of the junction. As the vortex elongates, bends and assumes a knee-like shape (at $\tau_4 = 0.6$ in figure~\ref{fig:breathing}), the particles follow increasingly complex trajectories and the reversal position shifts downstream. The same figure also reports, in cyan, the corresponding flow pathlines, which spiral around the primary vortex and leave the domain, in contrast to the trapped-particle trajectories.

\begin{figure}
	\centering
\includegraphics[scale=0.95]{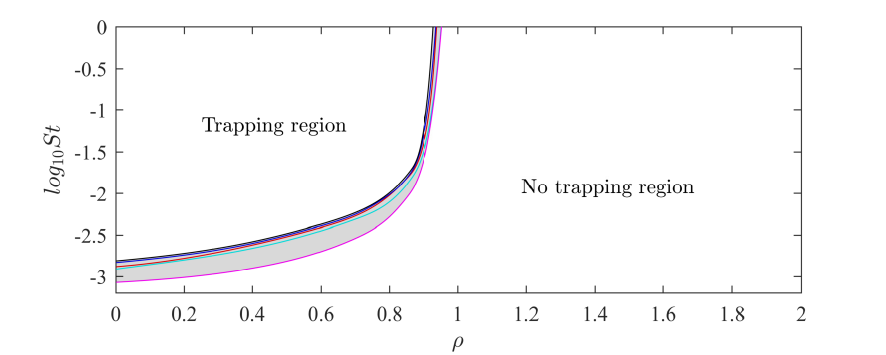}
\caption{Numerical threshold curves obtained solving the complete equation of motion \eqref{eqpm2} for different injection times at $Re=570$ and $Fr=0.24$. Red line, $\tau_1$; blue line, $\tau_2$; black line, $\tau_3$; cyan line, $\tau_4$; magenta line, $\tau_5$.}
	\label{fig:St_rho_570}
\end{figure}

As for the inner trajectories shown in figure~\ref{fig:trajectories_570}, they depend strongly on the injection time owing to the unsteady nature of the flow. When particles reach the primary vortex during its elongation phase, the corresponding trajectories reverse at a further downstream location than those of particles entering the vortical structure during the contraction phase. In the latter case, the rapid change in vortex shape gives rise to complex trajectories, with multiple loops forming near the stagnation point. By contrast, particles that travel further downstream follow more open trajectories, with motion inversion occurring at higher $z$ levels and without pronounced looping around the vortex core. These distinct behaviours are observed, for instance, for particles injected at $\tau_1$ and $\tau_3$, respectively.

Figure~\ref{fig:St_rho_570} depicts the threshold curves separating conditions leading to particle trapping from those allowing particles to exit the domain, in the $\rho-St$ plane, for different injection times. The curves are constructed by solving the complete equation of motion \eqref{eqpm2}. The dependence of the particle pathlines on the injection time also affects the threshold curves. For the various $\tau_i$ here considered, all curves fall in the light grey region of the figure. Particles reaching the vortical structure during its elongation period are more easily trapped due to the larger size of the recirculation zone, so that the corresponding trapping - no trapping threshold curves are located close to the outer boundary of the shaded area. In contrast, the particles entering the vortical structure during its contraction phase are trapped for lower values of $\rho$ and $St$ , thus defining the inner boundary of the light grey region. This finding is important from a practical viewpoint when designing a T-shaped channel flow capable of enhancing the particle capture.

\subsubsection*{Statistical analysis of the relative importance of the different forces along the trajectory}\label{subsubsec:forces_570}

The analysis is completed by the inspection of the magnitudes of the forces experienced by particles released at various injection times at two different locations. Because of the phase dependence of the particle dynamics on the breathing cycle, the results are discussed in terms of probability density function (PDF) and coefficient of variation (CV). The injection points considered are located at $(x,y,z)=(0.1,0.24,1)$ and $(0.20,0.24,1)$ and are indicated with star markers in the right panel of figure~\ref{fig:probability}. They correspond to a trapping probability of $P=1$ and $P=0.25$, respectively.

\begin{figure}
\captionsetup[subfigure]{labelformat=empty}
	\centering
  \subfloat[]{\includegraphics[scale=.9]{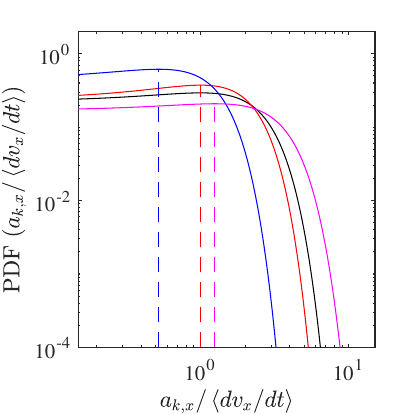}} \hspace{0.1cm}
	\subfloat[]{\includegraphics[scale=.9]{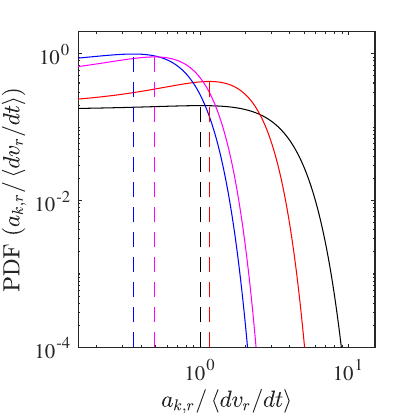}}\\
  \caption{Probability density functions of the particle acceleration (black line) and the different contributions due to pressure-gradient (red line), drag (blue line), and BB term (magenta line) along the axial (left panel) and radial (right panel) directions. The vertical lines represent the corresponding mean absolute values. $Re=570$ and $Fr=0.24$, $\rho=0.2$ and $St = 10^{-2}$. The injection point is at $x = 0.1$, $y = 0.24$ and $z = 1$.}	
	\label{fig:pdf1_x}
\end{figure}

\begin{figure}
\captionsetup[subfigure]{labelformat=empty}
	\centering
  \subfloat[]{\includegraphics[scale=.9]{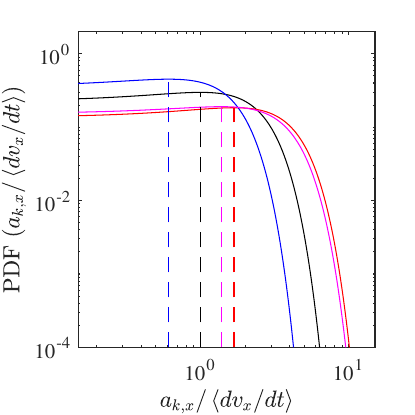}} \hspace{0.1cm}
	\subfloat[]{\includegraphics[scale=.9]{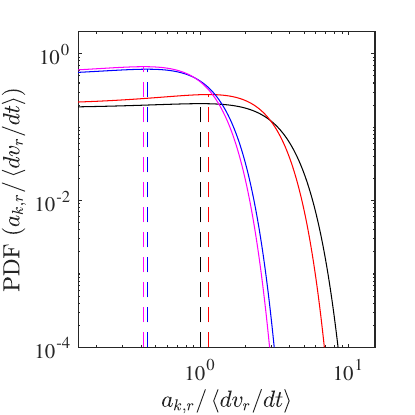}}\\
  \caption{Probability density functions of the particle acceleration (black line) and the different contributions due to pressure-gradient (red line), drag (blue line), and BB term (magenta line) along the axial (left panel) and radial (right panel) directions. The vertical lines represent the corresponding mean absolute values. $Re=570$ and $Fr=0.24$, $\rho=0.2$ and $St = 10^{-2}$. The injection point is $x = 0.2$, $y = 0.24$ and $z = 1$.}	
	\label{fig:pdf2_x}
\end{figure}
The PDFs of the magnitudes of the streamwise and radial force components, normalized by the average particle acceleration $\left\langle \partial v
/dt \right\rangle$, are considered. As done in \citet{Daitche2015567}, we examine the total particle acceleration and the different contributions due to the pressure-gradient, the drag, and the BB term. Note that the PDfs are calculated for various injection times, thus providing a global description of the force amplitudes experienced by the particle along its trajectory.
The results are shown in figures~\ref{fig:pdf1_x} and~\ref{fig:pdf2_x} for the two injection locations at $x=0.1$ and $x=0.20$, respectively. The left (respectively, right) panels refer to the force components in the axial (respectively, radial) direction. The vertical lines indicate the mean values of the different contributions to acceleration, due to pressure gradient ($\left\langle a_p\right\rangle$), viscous drag ($\left\langle a_d\right\rangle$) and memory force ($\left\langle a_{BB}\right\rangle$), relative to the particle acceleration $\left\langle \partial v/dt \right\rangle$. 

In the axial direction, when injecting the particle at $x=0.1$, the pressure-gradient and BB terms exhibit comparable average amplitudes, whereas the drag contribution is shifted toward lower values. This suggests that in the streamwise direction the particle behaviour is mainly influenced by both pressure-gradient and memory effects, while the drag term provides a minor average contribution. The figure also shows that the PDFs of $a_p$ and $a_{BB}$ have a longer tail than that of the drag, namely they produce more extreme events on the particle dynamics. Conversely, in the radial direction, $a_d$ and $a_{BB}$ are of the same order of magnitude, and the pressure gradient contribution has a longer tail and the separation of the tails is more pronounced. Physically this implies that radial particle motion is strongly influenced by local unsteady flow structures, that can then generate transverse forcing events. The extended tails observed in the radial direction further indicate that, although high-amplitude radial events are less frequent, they may still contribute significantly to particle redistribution across the flow.

For the injection location at $x=0.20$, the information given by the PDF analysis remains qualitatively similar. The differences in the average amplitude and tails extensions are representative of the lower trapping probability for a particle injected at such a location. As for the case $x=0.1$, in the axial direction the pressure gradient and the BB term have on the average larger amplitudes than the drag. Hence, the streamwise behaviour is again mainly driven by pressure gradient and memory effects, with the viscous drag providing a weaker average contribution. In the radial direction, both the drag and the BB term attain lower values than the prevailing pressure one. 

The PDFs presented above describe the overall distribution of the magnitude of the various acceleration contributions obtained for the ensemble of the events associated with all injection phases. To gain further information on the phase dependence of each statistical dataset, the investigation is corroborated by an analysis of the coefficient of variation ($CV_k(\tau_i)$) of the different contributions to the particle acceleration at every  injection time ($\tau_i)$,
\begin{equation}
CV_k(\tau_i)=
\frac{\sigma_{a_k}(\tau_i)}
{\left|\left\langle a_k \right\rangle_{\tau_i}\right|},
\label{eq:CV}
\end{equation}
where $\sigma_{a_k}(\tau_i)$ is the standard deviation of the $k$-th acceleration contribution when the particle is injected at  \(\tau_i\).

\begin{figure}
	\centering
\includegraphics[scale=0.95]{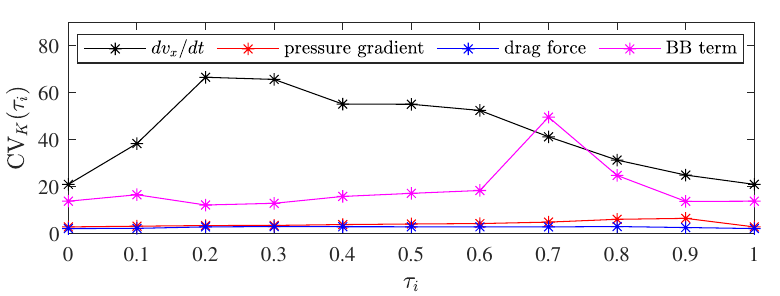}
\caption{Coefficients of variation for the different force contributions as functions of the injection time $\tau_i$ along the axial direction. $Re=570$ and $Fr=0.24$, $\rho=0.2$ and $St = 10^{-2}$. The injection point is $x = 0.1$, $y = 0.24$ and $z = 1$.}
	\label{fig:cv1_x}
\end{figure}

\begin{figure}
	\centering
\includegraphics[scale=0.95]{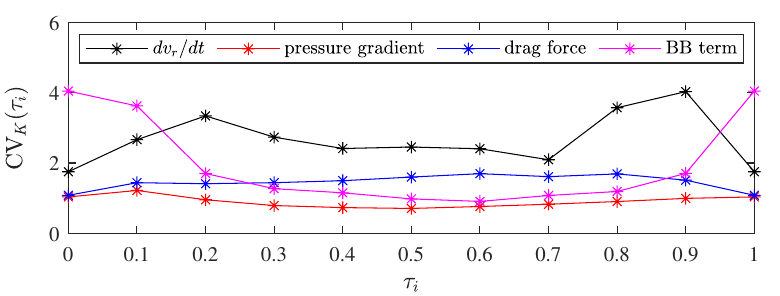}
\caption{Coefficients of variation for the different force contributions as functions of the injection time $\tau_i$ along the radial direction. $Re=570$ and $Fr=0.24$, $\rho=0.2$ and $St = 10^{-2}$. The injection point is $x = 0.1$, $y = 0.24$ and $z = 1$.}
	\label{fig:cv1_r}
\end{figure}

The resulting CVs, reported in figures~\ref{fig:cv1_x} and~\ref{fig:cv1_r} for particles injected at $x=0.1$, highlight a much higher variability of the forces acting on the particle in the axial direction than in the radial one. In particular, the axial values are up to one order of magnitude larger than the radial ones, indicating that the streamwise dynamics is much more sensitive to the injection phase. This is consistent with the trajectory dynamics reported in figure~\ref{fig:trajectories_570}: particles released at different phases reach substantially different downstream axial positions, whereas their radial locations remain comparatively close to each other. Therefore, the larger axial CV values provide a statistical signature of the phase-dependent variability of trajectory along the main flow direction, while the smaller radial values highlight a substantial phase-independent feature of the radial motion. This result, on the one hand, reinforces the major role played by the history force in the problem under consideration, significantly influencing BoA and particle capture On the other hand, it highlights that these characteristics strongly depend on the three-dimensional nature of the flow under study.
Since the results for $x = 0.2$ exhibit the same trend, they are not reported here.

\section{Conclusions}\label{sec:conclusions}

The present paper has investigated the motion of suspended inertial particles in a carrier fluid and the dynamics of the underlying vortical structures in a square microfluidic T-junction channel. It is known that the coherent structures forming in this type of flow strongly affect the particle distribution and may promote particle trapping around the T-region. Single-phase numerical simulations of the unperturbed base flow and integration of the particle force balance equation have been performed to study the particle motion along the trajectories for two Reynolds numbers, $Re$ equal to 500 and 570, respectively characterized by a steady vortex pair, and a helical oscillation (breathing mode) of the primary vortex pair due to a Hopf bifurcation. The base flow is investigated with direct numerical simulations (DNS) of the incompressible Navier-Stokes equations using the OpenFOAM solver.

The motion of a (spherical) particle is modelled by the momentum balance equation, assuming the particle subjected to the flow field acceleration (essentially the pressure gradient), the buoyancy force, the viscous drag due to the relative motion between the particle and the fluid, and the Basset-Boussinesq history force that takes into account the viscous diffusion of vorticity in the boundary layer surrounding the particle and that depends on the specific path described by the particle (memory effect). The motion equation has been numerically integrated by means of a suitable explicit numerical procedure of literature (\citet{Jaganathan2025135}) in MATLAB environment. 

As a major finding of the present work, we have shown that the BB term, that is ignored in most of the previous papers, has a significant influence on all flow conditions here analysed, characterized by flow $Re$ number and ($\rho$, $St$) pairs. Since the history force has been formulated considering the classic form of the kernel scaling in time as $t^{-1/2}$, after transforming the convolution integral into the time semiderivative, the force balance equation has been recast into a fractional ordinary differential equation, allowing us to express the solution in terms of Mittag-Leffler function. The presence of the fractional term slows down the time decay of the particle velocity during the initial transient period with respect to the rapid exponential decay of the classical solution (that accounts only for the local viscous drag).

After an \textit{a posteriori} reformulation of the motion equation in a cylindrical reference frame to project and analyse the contributions of the various forces relative to the vortex core, it has been noted that for the steady case $Re=500$ the BB contribution acts as an additional viscous force which, together with the drag, opposes the pressure-gradient force, making particle spiral trajectories wider in the radial direction and favouring its elongation in the axial direction. From a detailed analysis of the influence of the particle injection location on the trajectory, we have observed that Basset-Boussinesq force reduces the basin of attraction (BoA), indicating that memory effects promote the particle displacement away from the vortex core and weaken the trapping over the whole parameter range. In the $\rho-St$ plane, the BoA was identified by computing the threshold curve that separates the region where trapping occurs from the one where particles are not trapped. The result is that the threshold computed with the complete equation is entirely contained within that obtained without the memory term. This behaviour is in contrast with the finding of \citet{Kapoor2024} for a prototypal two-dimensional flow of a rotating pair of Lamb-Oseen vortices, whereas the configuration under investigation is dominated by strong three-dimensional effects.

At $Re=570$ (i.e., after the Hopf bifurcation), the flow exhibits a persistent oscillatory behaviour. SPOD based modal analysis has highlighted the occurrence of a fundamental frequency peak at Strouhal $S=0.34$, that is associated with the breathing mode of the primary vortex pair. We have interpreted this frequency in terms of an edge-tone phenomenon, as the characteristic frequency of the shedding of vortices generated at the edge of the recirculation bubble and issuing in a constant pressure ambient at the channel outlet section. In this unsteady case, for a given injection location, the influence of the injection time has also been considered to assess the dependence of the injection time on the phase of the vortex oscillation cycle. The resulting phase-dependent dynamics was found to influence both the shape and extent of BoA and was quantified in terms of the trapping probability for the entire injection inlet section. Because of the phase dependence of the particle dynamics on the breathing cycle, the investigation was finally corroborated by a suitable analysis to statistically evaluate the relative importance of the various forces along the trajectory. The magnitudes of the forces acting on the particle released at various injection times were examined in terms of probability density function and coefficient of variation. A significant result is that the axial particle behaviour is mainly driven by pressure gradient and memory effects, while a substantial phase-independent feature of the radial motion is observed. This result reinforces the major role played by the history force in the problem under consideration, and highlights that the key features of the trapping occurrence depend on the three-dimensional nature of the flow.
\backsection[Acknowledgements]{
Matteo Chiatto acknowledges the support of the Fulbright fellowship at Princeton University. Francesco Grasso acknowledges fundings as Global Scholar and as William Robert Jr. Keenan Distinguished Visiting Professor at Princeton University.}

\backsection[Declaration of interests]{The authors report no conflict of interest.}

\backsection[Authors ORCID]{\\
	\noindent 	Matteo Chiatto, 
	https://orcid.org/0000-0002-5080-7756,\\
	Luigi de Luca, 
	https://orcid.org/0000-0002-1638-0429.\\
    Howard A. Stone, 
	https://orcid.org/0000-0002-9670-0639.\\
	Francesco Grasso, 
	https://orcid.org/0000-0003-3826-6624.\\
}

\appendix

\section{Analytical solution of the fractional differential equation for particle velocity \label{sec:Mittag}}

After introducing the semiderivative  ${d^{1/2}\mathbf{v}}/{d t^{1/2}}$, given in equation \eqref{eq:fra_deriv}, the particle motion equation \eqref{eq:general_BB} can be recast in the form
\begin{equation}
\frac{d\mathbf{v}}{dt}  + a\frac{d^{1/2}\mathbf{v}}{d t^{1/2}}  +b\mathbf{v} = \mathbf{F}(t), 
\label{eq:symbol_frac}
\end{equation}
where $a=(3/\sqrt{2})(R/\sqrt{St})$ and $b=R/St$, with $\mathbf{F}(t)$ being the forcing.

By applying the Laplace transform method, the solution of the above equation may be expressed in terms of
Mittag--Leffler functions \citep{Gorenﬂo20201}, obtaining the equation \eqref{eq:formal_solution_BB}, that for convenience is reported below

\begin{equation}
\mathbf{v}(t)=H(t)\mathbf{v}_0+\int_{t_0}^{t}G(t-\tau)\,\mathbf{F}(\tau)\,d\tau,
\label{eq:formal_solution_in_A}
\end{equation}
where \(H(t)\mathbf{v}_0\) is the solution of the corresponding homogeneous problem (i.e. in the absence of forcing) depending on the initial condition $\mathbf{v}_0$.
$G(t)$ and $H(t)$ are related to the Mittag-Leffler function by the relationships:
\begin{equation}
G(t)=\frac{1}{\lambda_1-\lambda_2}\left[E_{1/2}(\lambda_1 t^{1/2})-E_{1/2} ( \lambda_2 t^{1/2}) \right],
\label{eq:G}
\end{equation}
and 
\begin{equation}
H(t)= \frac{1}{\lambda_1-\lambda_2}\left[\lambda_1E_{1/2}(\lambda_2 t^{1/2})-\lambda_2 E_{1/2} ( \lambda_1 t^{1/2}) \right],
\label{eq:H}
\end{equation}
where
\begin{equation}
\lambda_{1.2}= \frac{-a \pm \sqrt{a^2-4b}}{2}.
\label{eq:lambda}
\end{equation}

Note that in the present case a convenient expression for the Mittag-Leffler function is:
\begin{equation}
E_{1/2}(t)=e^{t^2}\textit{erfc}(
-t)
\label{eq:Mitt_Leff}
\end{equation}
where \textit{erfc} denotes the complementary error function.

As an example, to outline the key features of the solution, the equation \eqref{eq:symbol_frac} has been analysed in a simple case for an unknown scalar function $v(t)$, assuming $a=30$, $b=143$, $v_0$=1, $F(t)$=$\sin(\omega t)$, with $\omega=20$. The values of the coefficients $a$ and $b$ correspond to the physical parameters $\rho= 0.2$ and $St=10^{-2}$.

\begin{figure}
	\centering
\includegraphics[scale=0.95]{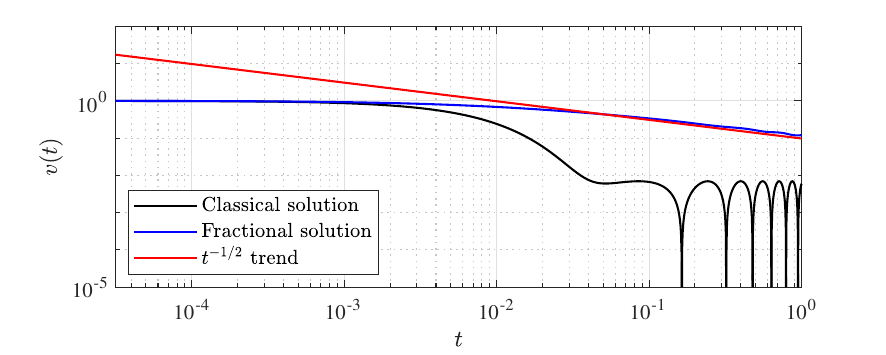}
\caption{Comparison between the solution of the complete  \textit{fractional} equation and the \textit{classical} one obtained by neglecting the memory term.}
	\label{fig:appendix_1}
\end{figure}

Figure~\ref{fig:appendix_1} depicts the comparison between the solution of the full \textit{fractional} equation and the \textit{classical} one obtained by neglecting the memory term (i.e., for $a=0$), and shows that the \textit{fractional} solution recovers the asymptotic trend $\approx t^{-1/2}$, as also found by \citet{Langlois2015} for the homogeneous problem. Observe that during the initial transient the presence of the fractional term slows down the time decay of $v(t)$, whereas the \textit{classical} solution exhibits a rapid exponential decay. This finding is confirmed for other values of the coefficients $a$ and $b$, corresponding to different pairs ($\rho$, $St$), and it is in some sense counter-intuitive since one would expect that an additional damping term (represented by the history effects) should strengthen the exponential decreasing trend. 

\bibliographystyle{jfm}
\bibliography{mybibfile}

\end{document}